\documentclass[twocolumn,showpacs,prd,superscriptaddress,aps,floatfix,longbibliography]{revtex4-1}
\usepackage{rotating}
\usepackage{amsmath}
\usepackage{amsfonts}
\usepackage{amssymb}
\usepackage{color}
\usepackage{graphicx}
\usepackage[caption=false]{subfig} % <=== "caption=false" added
\usepackage{courier}
\usepackage[active]{srcltx}
\usepackage[sort&compress]{natbib}

\newcommand{\qq}{{\bf q}}

\newcommand{\PP}{{\bf P}}

\newcommand{\kk}{{\bf k}}

\newcommand{\w}{\omega}

\newcommand{\be}{\begin{equation}}
\newcommand{\ee}{\end{equation}}
\newcommand{\ben}{\begin{equation*}}
\newcommand{\een}{\end{equation*}}
\newcommand{\bea}{\begin{eqnarray}}
\newcommand{\eea}{\end{eqnarray}}
\newcommand{\bean}{\begin{eqnarray*}}
\newcommand{\eean}{\end{eqnarray*}}

\def\efield{\boldsymbol{\cal E}}

\newcommand{\cnrs}{CNRS/Univ. Grenoble Alpes, Institut N\'eel, F-38042 Grenoble, France}
\newcommand{\cinam}{CNRS/Aix-Marseille Universit\'e, Centre Interdisciplinaire de Nanoscience de Marseille UMR 7325 Campus de Luminy, 13288 Marseille cedex 9, France}
\newcommand{\piim}{PIIM, Aix-Marseille Universit\'e}
\newcommand{\qub}{School of Mathematics and Physics, Queen's University Belfast, Belfast BT7 1NN, Northern Ireland, UK}
\newcommand{\etsf}{European Theoretical Spectroscopy Facilities (ETSF)}

\begin{document}
\title{Excitonic effects in third harmonic generation:\\ the case of carbon nanotubes and nanoribbons} 
%On the origin of carbon nanotubes non-linear response: \\correlation, saturation or intrinsic effects?}
\author{C. Attaccalite}
\affiliation{\cinam}
\affiliation{\cnrs}
\affiliation{\etsf}
\author{E. Cannuccia}
\affiliation{\piim}
\affiliation{\etsf}
\author{M. Gr\" uning} 
\affiliation{\qub}
\affiliation{\etsf}

\begin{abstract}
Linear and nonlinear optical properties of low dimensional nanostructures have attracted a large interest in the scientific community as tools to probe the strong confinement of the electrons and for possible applications in optoelectronic devices. In particular it has been shown that the linear optical response  of carbon nanotubes [Science 308, 838 (2005)] and graphene nanoribbons [Nat. Comm. 5, 4253 (2014)] is dominated by bounded electron-hole pairs, the excitons. The role of excitons in linear response has been widely studied, but still little is known on their effect on nonlinear susceptibilities. Using a recently developed methodology [Phys. Rev. B 88, 235113 (2013)] based on well-established ab-initio many-body perturbation theory approaches, we find that quasiparticle shifts and excitonic effects significantly modify the third-harmonic generation in carbon nanotubes and graphene nanoribbons. For both systems the net effect of many-body effects is to reduce the intensity of the main peak in the independent particle spectrum and redistribute the spectral weight among several excitonic resonances.
\end{abstract}

\maketitle
\section{Introduction}
Carbon nanotubes (CNTs) and graphene nanoribbons (GNRs) have remarkable electronic and optical properties due to their one-dimensional structure that combines solid-state characteristics with molecular dimensions. In these nanostructures  light  absorption produces strongly correlated electron-hole states in the form of excitons. Evidences of excitons  have  emerged  from  experimental studies  of  optical spectra and excited-state dynamics.\cite{Wang838,denk2014exciton}
The key role of excitons in the interpretation of the optical absorption of these materials has been confirmed by ab-initio computational studies\cite{PhysRevB.77.041404,PhysRevLett.92.077402} based on many-body perturbation theory (MBPT).\cite{strinati,aulbur1999quasiparticle} In fact the formation of strongly bounded excitons in GNRs has been theoretical predicted\cite{PhysRevB.77.041404} before the experimental measurement.\cite{denk2014exciton} 
In recent years, also the nonlinear optical response of these low dimensional structures has attracted a large attention, both from a fundamental and an applicative point of view. In particular due to their strong nonlinear response, one dimensional nanostructures find application as nanoantennas and optical switches.\cite{chen2002ultrafast,tatsuura2003semiconductor}\\
Experimentally, the absolute measure of nonlinear optical responses of these nanostructures is, however, not straightforward.\footnote{See for example Y. R. Shen ``The Principles of Nonlinear Optics'', Wiley Interscience (2003)}
The first non-zero non-linear response function in carbon nanoribbon and non-chiral nanotubes is  the third-order susceptibility.\\
In CNTs, only few measurements on third-order nonlinear susceptibility  have been reported so far and most of them on the $\chi^{(3)}(-\omega; \omega, -\omega, \omega)$\cite{maeda2005large,liu1999third,Ichida20091794,lauret2004third}, responsible for the intensity dependence of the refractive index. The one study on the third-harmonic-generation, $\chi^{(3)}(-3\omega; \omega, \omega, \omega)$,\cite{stanciu2002experimental,nemilentsau2006third} explored the nonperturbative regime. Regarding the GNRs, to our knowledge there are no available experimental measures of the THG though recently measurements of the THG in graphene have been obtained\cite{kumar2013third,hong2013optical,SLGTHG}. Interestingly these measurements found that graphene's THG is of the order of $10^{-15}-10^{-16} m^2/V^2$ ($10^{-7}-10^{-8}$ esu), thus comparable to resonant THG in bulk materials.

Theoretically, state-of-the-art calculations of nonlinear optical responses of periodic systems neglect excitonic effects that are deemed essential to describe optical properties in these carbon nanostructures.
For CNTs theoretical studies on THG---which mostly focused on radius and chirality dependence---have been performed at the independent particle approximation\cite{margulis1998theoretical,Zhang2006101,rezania}. 
The only work which includes many-body effects to our knowledge is from Lacivita and coworkers~\cite{lacivita2016longitudinal} that computed the \emph{static} second hyperpolarizability of CNTs within the coupled Kohn-Sham (KS) equations formalism.\footnote{Note that none of the mentioned studied included temperature effects, except Ref.~\onlinecite{rezania} which includes temperature effects by means of Green's function theory.} %The focus of these studies have been mostly the dependence of the THG on the CNTs radius and chirality.
For GNRs the nonlinear properties have been addressed in few studies\cite{wang2016first,C3TC31847H,karamanis2013second,ahmadia2011theoretical} that focused mainly on the possibility of enhancing the nonlinear response by engineering the ribbon edges.\cite{C3TC31847H,karamanis2013second} The works in Refs.\onlinecite{C3TC31847H,karamanis2013second} are based on time-dependent density functional theory with an hybrid approximation for the exchange-correlation functional, which approximately accounts for excitonic effects. Those studies however have been performed on \emph{finite length} GNRs.   

%The main focus of  these studies has been the dependence of the nonlinear optical properties of nanotubes on their electronic structure and on geometrical parameters. For instance Margulis et al.\cite{margulis1998theoretical} found that the magnitude of the THG grows with the CNT radius $R$ as $R^4$  and that in general optical nonlinearites are larger in small gap nanotubes. These results have been later confirmed by other calculations.\cite{Zhang2006101,lacivita2016longitudinal,rezania}

%\MG{The role of excitons in the nonlinear response is still unclear. Are the excitons enhancing the nonlinearity or is it the inverse? Getting an answer from the experiments is quite difficult and the major part of the theoretical studies, due to the complexity of the nonlinear response, have been limited to the independent particle approximation (IPA).}

\begin{figure}[t]
\vspace{1.0cm}
\includegraphics[width=.5\textwidth]{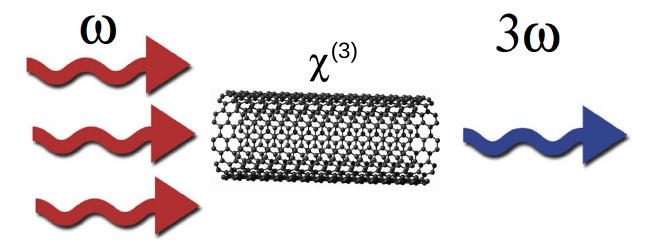}
\caption{[Color online] Schematic description of the THG process: three photons of frequency $\omega$ are destroyed and one photon of frequency $3\omega$ is created, i.e. the system responds at frequency $3\omega$ to an applied field at frequency $\omega$..\label{tubex3}}
\end{figure}

Here we present a theoretical/computational combined study of THG in CNTs and GNRs (Fig.~\ref{tubex3}) that includes excitonic effects. Specifically, we use an ab-initio approach based on many-body perturbation theory (MBPT) which has been shown to provide accurate results for both linear and nonlinear optical properties of periodic systems.\cite{strinati,aulbur1999quasiparticle,attaccalite2015strong,gruning2014}
With respect to the works mentioned above which include correlation beyond the independent particle approximation, our study addresses \emph{infinite periodic} nanostructures (at difference with Ref.\onlinecite{C3TC31847H,karamanis2013second}) and \emph{frequency dependent} third order response (at difference with Ref.~\onlinecite{lacivita2016longitudinal}). Furthermore our MBPT approach does not depend on any semi-empirical parameter at difference with hybrid density-functional based approaches used for previous works. In the latter, the semi-empirical mixing parameter eventually determines both the fundamental band gap and the amount of dielectric screening.  

The purpose of this work is threefold: first, to provide an accurate theoretical estimate of the THG in a small semiconducting CNT and GNR;  second and more importantly---as we can switch on and off excitonic and many-body effects in the (effective) Hamiltonian of the electron system---to evaluate how those effects affect the THG of the material; finally, to provide a benchmark to assess the reliability and accuracy of calculations at the independent-particle level which neglects these effects. The paper is organized as follows: in section \ref{computational} we summarize the computational methods employed in the calculation of the electronic structure and nonlinear response; in section \ref{sec:results} we present results for both the CNTs and GNRs. We consider here only centrosymmetric systems for which the THG is the first non-negligible nonlinear response.\footnote{A centrosymmetric system has an inversion center and thus the second-harmonic generation is zero within the electric dipole approximation. When multipole corrections are considered the second-harmonic generation is nonzero, but still very small.}

\section{Computational Methods}
%%%%%%%%%%%%%%%%%%%%%%%%%%%%%%%%%%%%%%%%%%%%%%%%%%%%%%%%%%%%%%%%%%%%%%%%%%%%%%%%%%%%%%%%%%%%%%%%%%%%%%%%%%%%%%%%%%%%%%%%%%%%%%%%%%%%%%%%
\label{computational}
The nanotubes and nanoribbons atomic structures have been generated from the ideal graphene with bonding length of $1.421$\AA. Subsequently atomic positions have been optimized by means of Density Functional Theory (DFT), using the local density approximation for the exchange correlation functional.\cite{PhysRevB.23.5048,PhysRevB.45.13244}

All DFT calculations have been performed with the Quantum Espresso code\cite{pwscf} where the wave functions are expanded in plane waves with a cutoff of $60$~Ry and the effect of core electrons is simulated by norm-conserving pseudopotentials.\cite{troullier} We used a $1 \times 1 \times 22$ $\kk$-points grid to converge the density in the CNT and GNR (both structures are oriented along the $z$ axis).
Valence and conduction orbitals that enter in the Green's function theory are obtained from the diagonalization of the KS  eigensystem. 
The KS eigensolutions $\{\varepsilon_{n\kk}; |n\kk\rangle\}$ correspond to the energies and Bloch wave functions (with $\kk$ the crystal wave vector and $n$ the band index) of the independent particle system that reproduces the electronic density of the system under study. 
 \begin{figure*}
\centering
% \subfloat{\label{fig:cnt100}\includegraphics[width=0.48\textwidth]{NEWRESULTS/CNT100/BANDS/bands_and_dos.pdf}}%width=3.2cm
 \subfloat{\label{fig:cnt100}\includegraphics[width=0.48\textwidth]{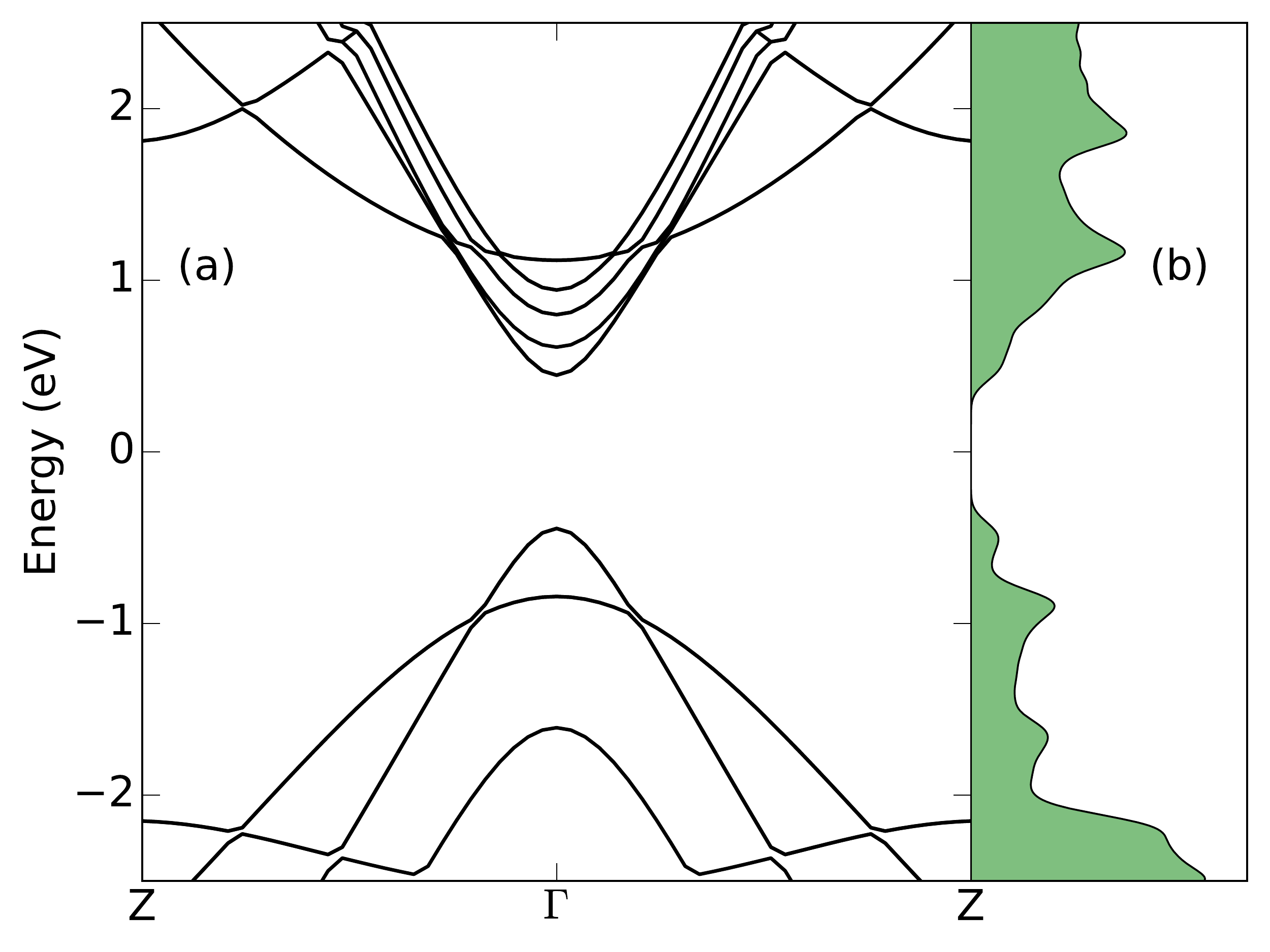}}%width=3.2cm
 \quad
%\subfloat{\label{fig:angr9}\includegraphics[width=0.48\textwidth]{NEWRESULTS/ANGR9/BANDSndDOS/bands_and_dos.pdf}}
\subfloat{\label{fig:angr9}\includegraphics[width=0.48\textwidth]{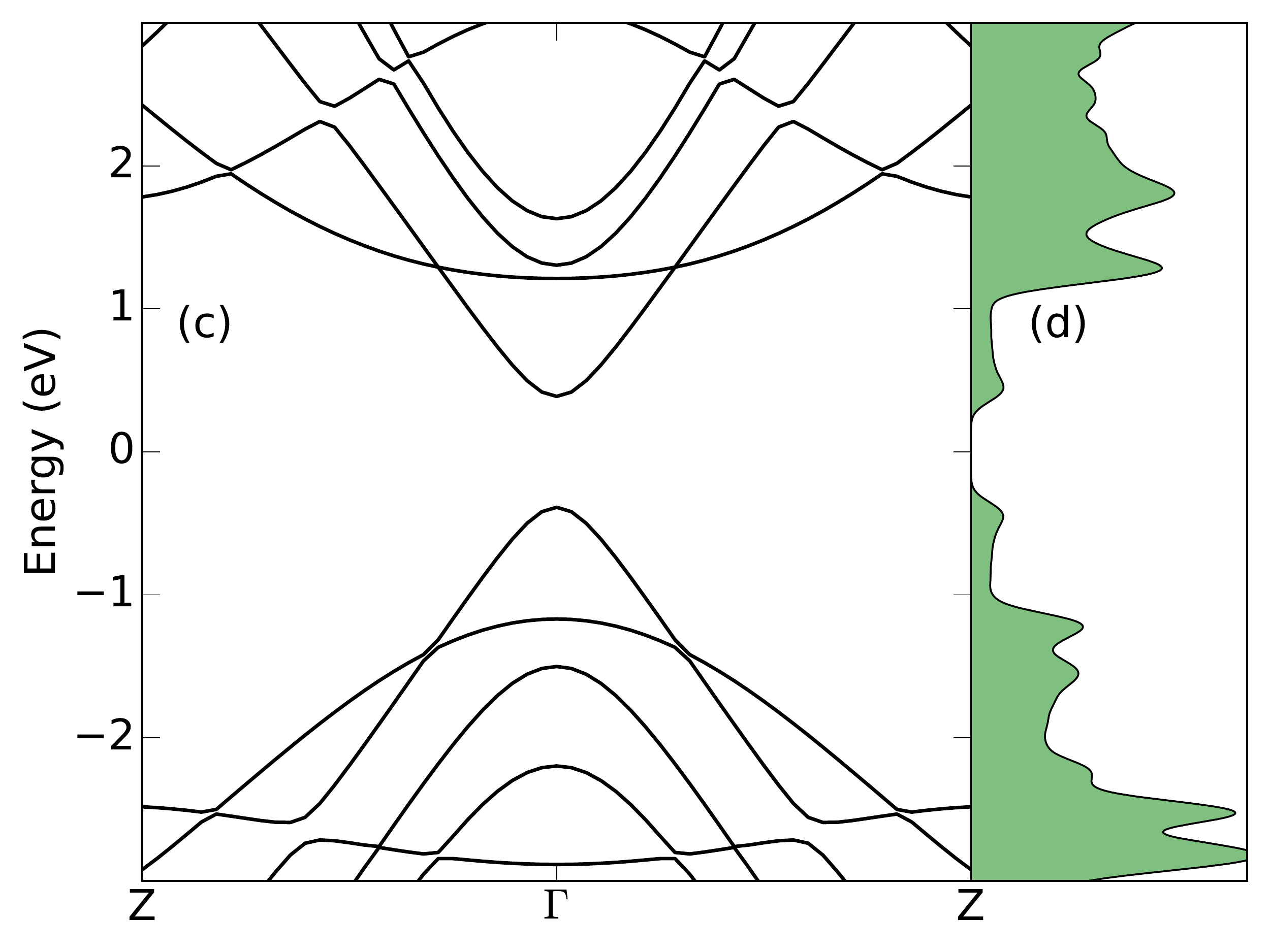}}
	 \caption{Band structure and density of states for the CNT(10,0)  [panels a) and b)] and the  9-AGNR [panels c) and d)]. Notice that the band structures and denities of states are aligned in such a way to have the same y-axis. The density of state is normalized in such a way that its integral on the occupaied bands give the total number of electrons.}
\label{fig:bands}
\end{figure*}
In order to simulate isolated nanotubes and nanoribbons we used a supercell approach with a tube-tube distance larger than the nanotube diameter and a distance between the ribbons of 16 a.u. in the perpendicular direction and larger than the ribbon size in the plane.  
\subsection{Quasiparticle band structure}
Starting from the KS eigenvalues and eigenfunctions, we obtained the quasiparticle (QP) band structure by means of MBPT within the so-called $GW$ approximation. Specifically, we use non-self consistent $GW$, often denoted as $G_0W_0$, in which the screened Coulomb potential, $W$, and the Green's function, $G$, are built from the KS eigensolutions  $\{\varepsilon_{n\kk}; |n\kk\rangle \}$  and the quasiparticle energies are obtained from:
\be
\varepsilon^{\text{QP}}_{n\kk} = \varepsilon_{n\kk} + Z_{n\kk} \Delta \Sigma_{n\kk}(\varepsilon_{n\kk}).
\label{per}
\ee
In Eq.~\ref{per}
$$Z_{n\kk} = [1-\partial \Delta \Sigma_{n\kk}(\omega) / \partial \omega |_{\omega=\varepsilon_{n\kk}}]^{-1},$$
is  the renormalization factor and
$$\Delta \Sigma_{n\kk} \equiv \langle n\kk |\Delta \hat \Sigma |n\kk \rangle,$$ 
where
$$\Delta \hat\Sigma = \hat\Sigma-\hat V^{\text{xc}},$$ 
is the difference between  $\Sigma =GW$, the $GW$ self-energy, and $V^{\text{xc}}$, the exchange-correlation functional used in the KS calculation.\cite{aulbur1999quasiparticle} The screened Coulomb potential $W$ has been evaluated within the random-phase approximation. In the $GW$ approach we used the Godby-Needs plasmon-pole model to approximate the dynamical behavior of $W$,\cite{godby} while in the Bethe-Salpeter equation (BSE) framework described in the next subsection we use the static approximation.\cite{strinati} In the $GW$ (and BSE) calculations we used a truncated Coulomb potential to reduce the interaction between the periodic replica.\cite{rozzi2006exact} 
The Green's function and the self-energy that appear in the $GW$ calculations are expanded in the basis of the KS eigensolutions: we used a 40 $\kk$-points and 320 bands for the CNT and 60 $\kk$-points and 200 bands for the GNR.
\subsection{Linear and nonlinear response functions}
\label{sec:resp}
Linear and nonlinear optical properties are obtained by means of a real-time implementation of the BSE.\cite{strinati,schmidt2003efficient} We used an effective Hamiltonian that includes electron-hole interaction through a screened exchange interaction and the coupling between electrons and the external field is described by means of modern theory of polarization.\cite{souza_prb,nloptics2013}  This formulation allows us to correctly describe  response functions beyond the linear one.\cite{nloptics2013}

Specifically we solve a set of coupled one-particle effective time-dependent Schr\"odinger equations: 
\bea
i\hbar  \frac{d}{dt}| v_{m\kk} \rangle &=& \left( H^{\text{sys}}_{\kk} +i \efield \cdot \tilde \partial_\kk\right) |v_{m\kk} \rangle \label{tdbse_shf},
\eea
where $| v_{m\kk} \rangle$ is the periodic part of the Bloch states that determine the system polarization~\cite{souza_prb} as discussed below.
In the r.h.s. of Eq.~\ref{tdbse_shf}, $ H^{\text{sys}}_{\mathbf k}$ is the system Hamiltonian---which is discussed later in this Section; the second term, $\efield \cdot \tilde \partial_\kk$, describes the coupling with the external field $\efield$ in the dipole approximation. As we  imposed Born-von K\'arm\'an periodic boundary conditions, the coupling takes the form of a $\kk$-derivative operator $\tilde \partial_\kk$. The tilde indicates that the operator is ``gauge covariant'' and guarantees that the solutions of Eq.~\ref{tdbse_shf} are invariant under unitary rotations among occupied states at $\kk$ (see Ref.~\onlinecite{souza_prb} for a discussion on this point).  
 
From $| v_{m\kk} \rangle$, the time-dependent polarization of the system  $P_\parallel$  along the lattice vector $\mathbf a$ is calculated as
 \begin{equation}
 P_\parallel = -\frac{ef |\mathbf a| }{2 \pi \Omega_c} \mbox{Im log} \prod_{\kk}^{N_{\kk}-1}\ \mbox{det} S\left(\kk , \kk + \mathbf q\right), \label{berryP} 
 \end{equation}
where $S(\kk , \kk + \mathbf q) $ is the overlap matrix between the valence states $|v_{n\kk}\rangle$ and $|v_{m\kk + \qq}\rangle$. Furthermore, $\Omega_c$ is the unit cell volume,  $f$ is the spin degeneracy, $N_{\kk}$ is the number of $\kk$ points along the polarization direction, and $\mathbf q = 2\pi/(N_{\kk} {\mathbf a})$.
Finally, the third harmonic coefficient is extracted from the expansion of the polarization in the laser field $\efield$ power series:
\be
\PP=\chi^{(1)} \efield + \chi^{(2)} \efield^2 + \chi^{(3)} \efield^3 + ...
\ee
as detailed in Ref.~\onlinecite{nloptics2013}. 

Notice that our approach to calculate  nonlinear susceptibilities does not work for metallic systems or systems with a very small gap, since it uses the  polarization [Eq.\ref{berryP}] as fundamental quantity. For metallic systems other approaches based on the electron current-density should be used instead.\cite{springborg}

\begin{figure}[ht]
\centering
\includegraphics[width=.5\textwidth]{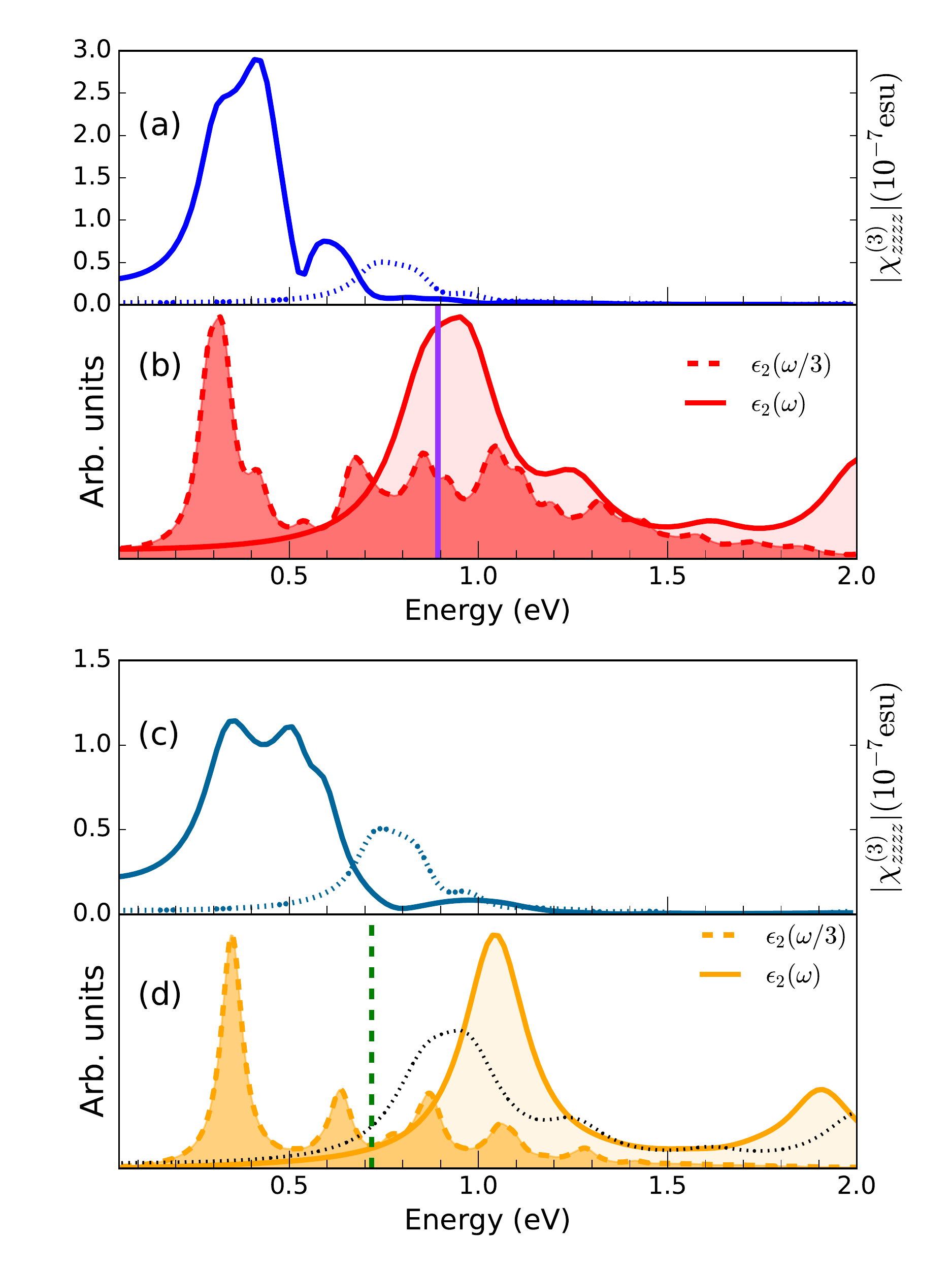}
	\caption{[Color online] THG intensity $|\chi^{(3)}_{zzzz}(\omega)|$ and optical absorption in the longitudinal direction ($z$-axis) for the CNT(10,0). Panel (a): THG from the IP model (continuous line) and the QP model (dotted line) for the Hamiltonian; panel (b): the imaginary part of the dielectric function at $\omega$ (continuous line) and $\omega/3$ (dashed line) from the IP model for the Hamiltonian; panel (c): THG within the GW+BSE model (continuous line) and QP model (dotted line) Hamiltonian;  panel (d): the imaginary part of the dielectric function at $\omega$ (continuous line) and  $\omega/3$ (dashed line) at the same level of approximation, IP results at $\omega$ (black dotted line). The vertical violet line represents the KS fundamental gap and the dashed line represents $1/3$ of the $GW$ fundamental gap.\label{figCNT100}}
\end{figure}

In Eq.~\ref{tdbse_shf} the model Hamiltonian chosen for  $ H^{\text{sys}}_{\mathbf k}$ determines the level of approximation in the description of correlation effects in the linear and nonlinear spectra. In this work we use different models for the system Hamiltonian: {\it(i)} the independent-particle (IP) model,
\be
   H^{\text{IP}}_{\mathbf k} \equiv H^{\text{KS}}_\kk,
\label{rt-ip}
\ee
where $H^{\text{KS}}_\kk$ is the unperturbed KS Hamiltonian and we consider the KS system  as a system of independent particles. The QP model  {\it(ii)},
\be
   H^{\text{QP}}_{\mathbf k} \equiv H^{\text{KS}}_\kk + \Delta H_\kk.
\label{rt-qp}
\ee
where a scissor operator $\Delta H_\kk$, estimated from the MBPT [Eq.~\ref{per}], has been applied to the KS eigenvalues.
%and {\it(iii)} the RPA model,
%\be
%   H^{\text{RPA}}_{\mathbf k} \equiv H^{\text{KS}}_\kk + + V_h(\mathbf r)[\Delta\rho],
%\label{rt-rpa}
%\ee
Finally  {\it(iii)} the full $GW$+BSE model reads,
\be
   H^{GW+\text{BSE}}_{\mathbf k} \equiv H^{\text{KS}}_\kk + \Delta H_\kk + V_h(\mathbf r)[\Delta\rho] + \Sigma_{\text{SEX}}[\Delta \gamma],
\label{rt-bse}
\ee
where $$\Delta \rho \equiv \rho(\mathbf r;t)-\rho(\mathbf r;t=0)$$ is the variation of the electronic density  and $$\Delta \gamma \equiv \gamma(\mathbf r,\mathbf r';t) - \gamma(\mathbf r,\mathbf r';t=0)$$ is the variation of the density matrix induced by the external field $\efield$.

In Eq.~\eqref{rt-bse} the term $V_h(\mathbf r)[\Delta\rho]$ is the time-dependent Hartree potential\cite{attaccalite} and is responsible for the local-field effects\cite{PhysRev.126.413} originating from system inhomogeneities. In the same equation, $\Sigma_{\text{SEX}}$ is the screened-exchange self-energy that accounts for the electron-hole interaction,\cite{strinati} and is given by the convolution between the screened interaction $W$ and $\Delta \gamma$. In the limit of small perturbation Eq.~\ref{rt-bse} reproduces the optical absorption calculated with the standard $GW$ + BSE approach\cite{strinati}, as shown both analytically and numerically in Ref.~\onlinecite{attaccalite}.

We calculate the $|\chi^{(3)}_{zzzz}(-3\w;\w;\w;\w)|$, i.e. the magnitude of the third order nonlinear susceptibility at a frequency of $3\w$ along the $z$ axis (the orientation axis of nanotubes and the ribbons) when a monochromatic  electric field of frequency $\w$ is applied along $z$. 
This quantity is obtained by integrating Eq.~\eqref{tdbse_shf} numerically for a time interval of $120$~fs using the numerical approach described in Ref.~\onlinecite{souza_prb} (originally taken from Ref.~\onlinecite{koonin90}) with a time step of $\Delta t = 0.01$~fs, which guarantees for numerically stable and sufficiently accurate simulations. 
$|\chi^{(3)}_{zzzz}|$ is finally extracted from the total polarization after 100~fs as described in Ref.~\onlinecite{nloptics2013}. A dephasing term with a time $\tau=8.78$~fs is introduced in Eq.~\eqref{tdbse_shf} in order to simulate a finite broadening of about $0.15$~eV.\cite{nloptics2013}
Finally note that our calculations provide the volume third-order nonlinear susceptibility, but
because of the use of the supercell, two of the dimensions are not physical. We then rescale by the effective physical dimensions of the systems: for both systems we use $0.335$~nm as effective thickness and for the GNRs we consider as effective width the width of the ribbon plus $0.2$~nm. 
%%%%%%%%%%%%%%%%%%%55

\section{Results}
\label{sec:results}
\subsection{Carbon nanotube}
A single-walled CNT is formed by  rolling  a  sheet  of  graphene  into  a  cylinder  along  an $(m,n)$ lattice vector in the graphene plane. 
These two indexes determine the diameter and chirality, which are  key  parameters  of  a  nanotube.  Depending  on  the chirality (the chiral angle between hexagons and the main CNT axis), nanotubes can be either metals or semiconductors, with band gaps that may vary between few meV to the eV, even if they have nearly  identical  diameters.\cite{smalley2003carbon} We consider here semiconducting zig-zag CNTs only.

We analyze the effect of correlation in THG within the $GW$+BSE approach as detailed in Sec.~\ref{sec:resp}. Due to the computational cost we limit the calculation to the (10,0) CNT. We expect the analysis to be valid for larger zig-zag CNTs.

In Fig.~\ref{figCNT100} we report the THG magnitude $|\chi^{(3)}_{zzzz}(\omega)|$ at different levels of approximation [panels (a) and (c)] and the corresponding optical absorption [panels (b) and (d)] so to identify the resonances in the THG spectrum from the comparison.

At the IP level [panel (b) of Fig.~\ref{figCNT100}] the absorption spectrum is dominated by a large peak at about $0.9$~eV corresponding to the transition between the highest valence and lowest conduction bands at $\Gamma$ [see band structure,  panel (a) of Fig.~\ref{fig:bands}]. Transitions between second-highest valence and the lowest conductions bands give rise to the shoulder around $1.3$~eV, that is visible also in the THG spectra. Peaks at higher energies are originated from transitions from the third-highest valence bands.

Turning to the THG, at the level of the IP model [panel (a) of Fig.~\ref{figCNT100}] we observe at about $0.3$~eV and $0.4$~eV three photon-resonances corresponding to the main peaks in the absorption spectrum. Notice that in the THG the intensity of the peaks is reversed respect to the linear optics. At $0.6$~eV we found a smaller three photon-resonance peak that stems from the transition from the third-highest valence band. One-photon resonance peaks are barely visible on this scale. Our results are consistent with those obtained by Nemilentsau et al.\cite{Nemilentsau20062246} and Xu et al.~\cite{xu2004third}. With respect to Nemilentsau and coworkers our calculated linear and nonlinear response functions have a richer structure as we use a full \emph{ab-initio} band structure rather than the two-bands model employed in their work. 

Inclusion of QP corrections rigidly blue-shifts the absorption spectrum by approximately $1.6$~eV (not shown). When one includes as well the electron-hole interaction [$GW$+BSE model in Fig.~\ref{figCNT100} panel (d)] the spectrum is red-shifted with respect to the QP model by $1.5$~eV (excitonic binding energy). As a result the resonance peak is blue-shifted by about $0.1$~eV with respect to the IP level. More remarkably the spectrum is dominated by a strong excitonic peak at $1.1$~eV which almost doubles in intensity the van Hove singularity in the IP spectrum. 

In the THG spectrum the inclusion of QP corrections blue-shifts it by about $0.5$~eV and reduces the spectral intensity by almost one order of magnitude due to the sum rules constraints.\cite{saarinen2002sum} Similarly to what is observed for the absorption spectrum the inclusion of electron-hole interaction [panel (c) of Fig.~\ref{figCNT100}] red-shifts the spectrum and enhances the spectral intensity with respect to the QP model. The spectral enhancement is however not uniform: the  intensity of the main peak is doubled---though it is still about 1/3 of the main peak intensity in the IP model. On the other hand peaks at higher frequencies aquire a comparable intensity as the main peak. The one-photon resonance with the main peak remains very weak.

Notice that the $GW$+BSE calculations include also the so-called local field effects, namely the response of the time-dependent Hartree term to the external field. These corrections are large for inhomegenous systems, such as isolated molecules or localized orbitals and are exactly zero for an homogeneous electron gas. In the linear response of one-dimensional systems (e.g. CNTs) local field effects are negligible along the periodic direction and large for the perpendicular directions. We found that in the periodic direction the local fields effects are negligible for the THG as well.  

\begin{figure*}
\centering
% \subfloat{\label{fig:cnts}\includegraphics[width=0.48\textwidth]{NEWRESULTS/CNTs.pdf}}
 \subfloat{\label{fig:cnts}\includegraphics[width=0.48\textwidth]{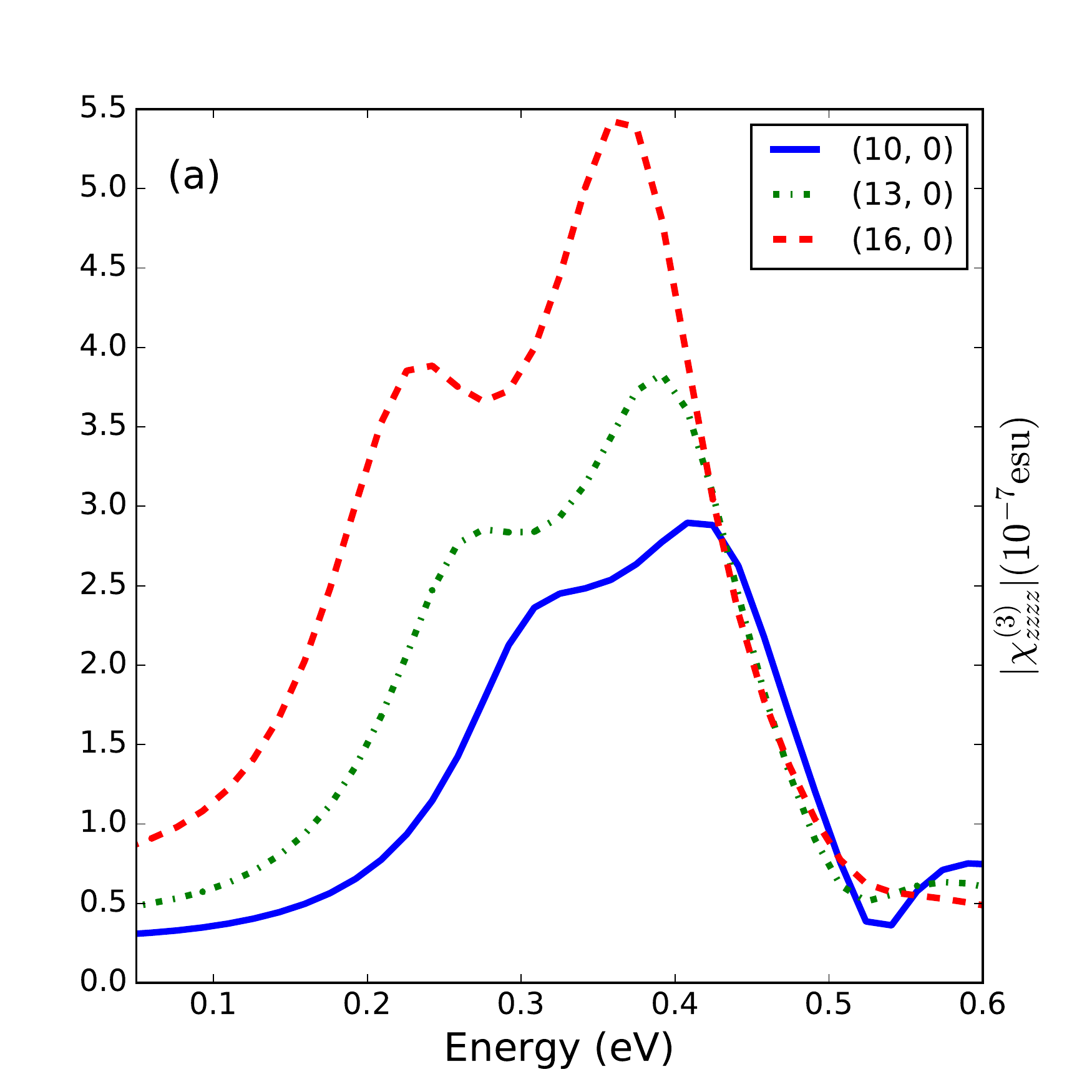}}
 \quad
%\subfloat{\label{fig:gnrs}\includegraphics[width=0.48\textwidth]{NEWRESULTS/GNRs.pdf}}
\subfloat{\label{fig:gnrs}\includegraphics[width=0.48\textwidth]{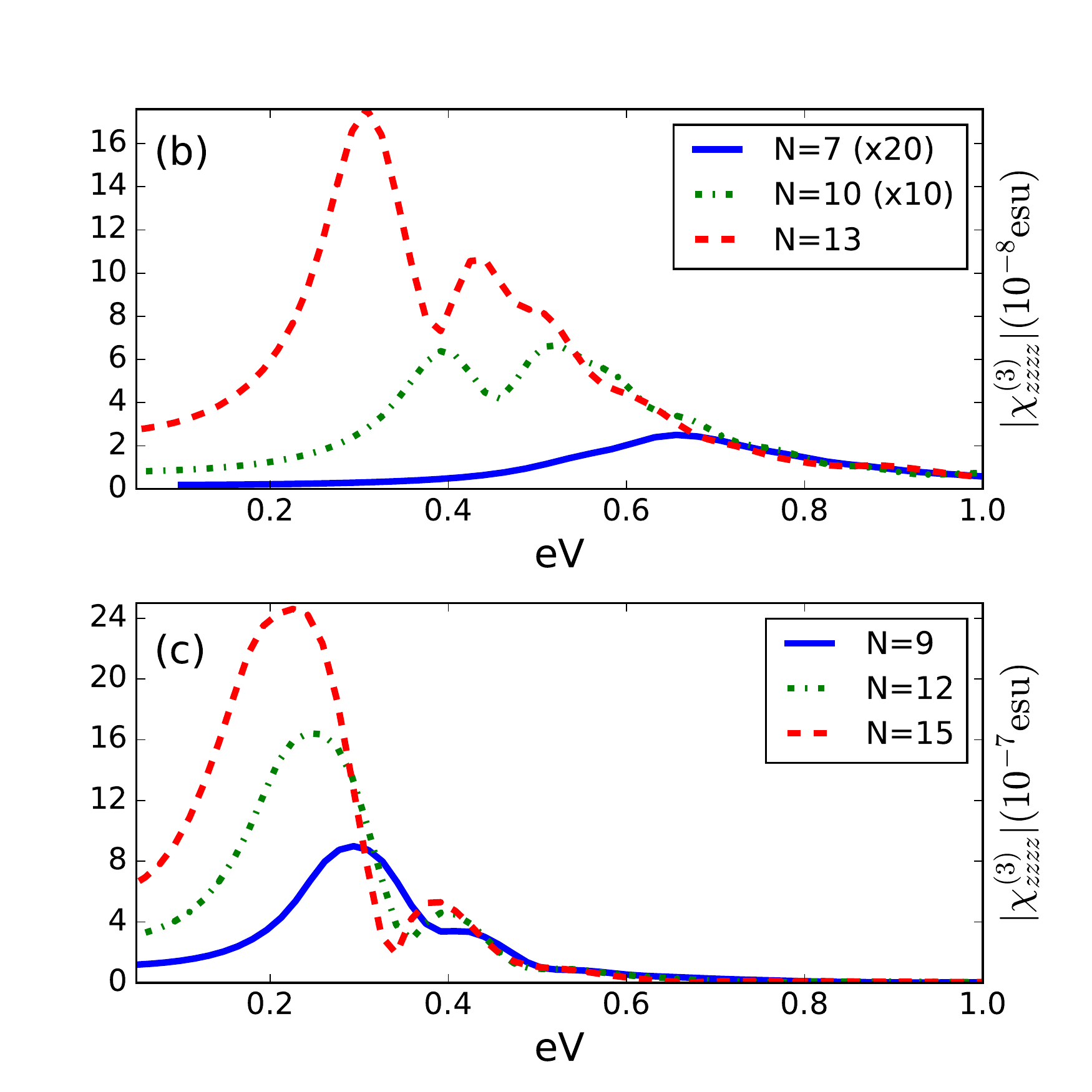}}
 \caption{THG response in GNRs and CNTs at the IP level. Panel (a) shows the THG for zig-zag CNTs of increasing size. Panel (b) shows the THG for the $N=3p+1$ GNR family and in panel (c) for the $N=3p$ family.}
\label{fig:manythg}
\end{figure*}

Finally, at the IP level only we study how the THG varies with the size of the CNT [Fig.\ref{fig:manythg} panel (a)]. Consistently with what found previously\cite{margulis1998theoretical,Zhang2006101,xu2004third} the THG increases superlinearly with the CNT radius. The resonant peaks shift to lower energy as a consequence of the gap shrinking for larger CNTs. The two peaks however do not shift of the same amount and they become more apart in larger CNTs.
We found a factor of about 1.4 between the peaks of the (13,0) and (16,0) CNTs in agreement with Refs.~\cite{margulis1998theoretical,xu2004third} which predicted a factor of about 1.5. The peaks magnitude is in agreement with that of Margulis and co-workers~\cite{margulis1998theoretical} while other works have reported larger magnitude especially Xu and coworkers whose reported THG at resonances is larger by almost two orders of magnitude. Regarding the spectral shape, it is clear that the two-band model employed in previous works is not sufficient and at least the second-highest valence band needs to be included. 

%\MG{This effect has been already observed in previous studies and in general it has been shown that the THG response increase for small gap nanotubes. From our results we get a THG spectra with an intensity comparable to previous published results that presents a more complex structure due to the use of a full band structure and not a simplified two-band model. CHANGE THIS}

\subsection{Armchair nanoribbon}
Armchair nanoribbons are divided in three distinct families depending on the ribbon width, namely $N = 3p$, $N = 3p + 1$, and $N = 3p + 2$, N being the number of dimer lines along the width, with $p$ a positive integer.  Within each family, the fundamental band gap decreases with increasing ribbon width.\cite{son2006energy} We consider here the semiconducting $N = 3p$, $N = 3p + 1$ families only.

\begin{figure}[ht]
\centering
\includegraphics[width=.5\textwidth]{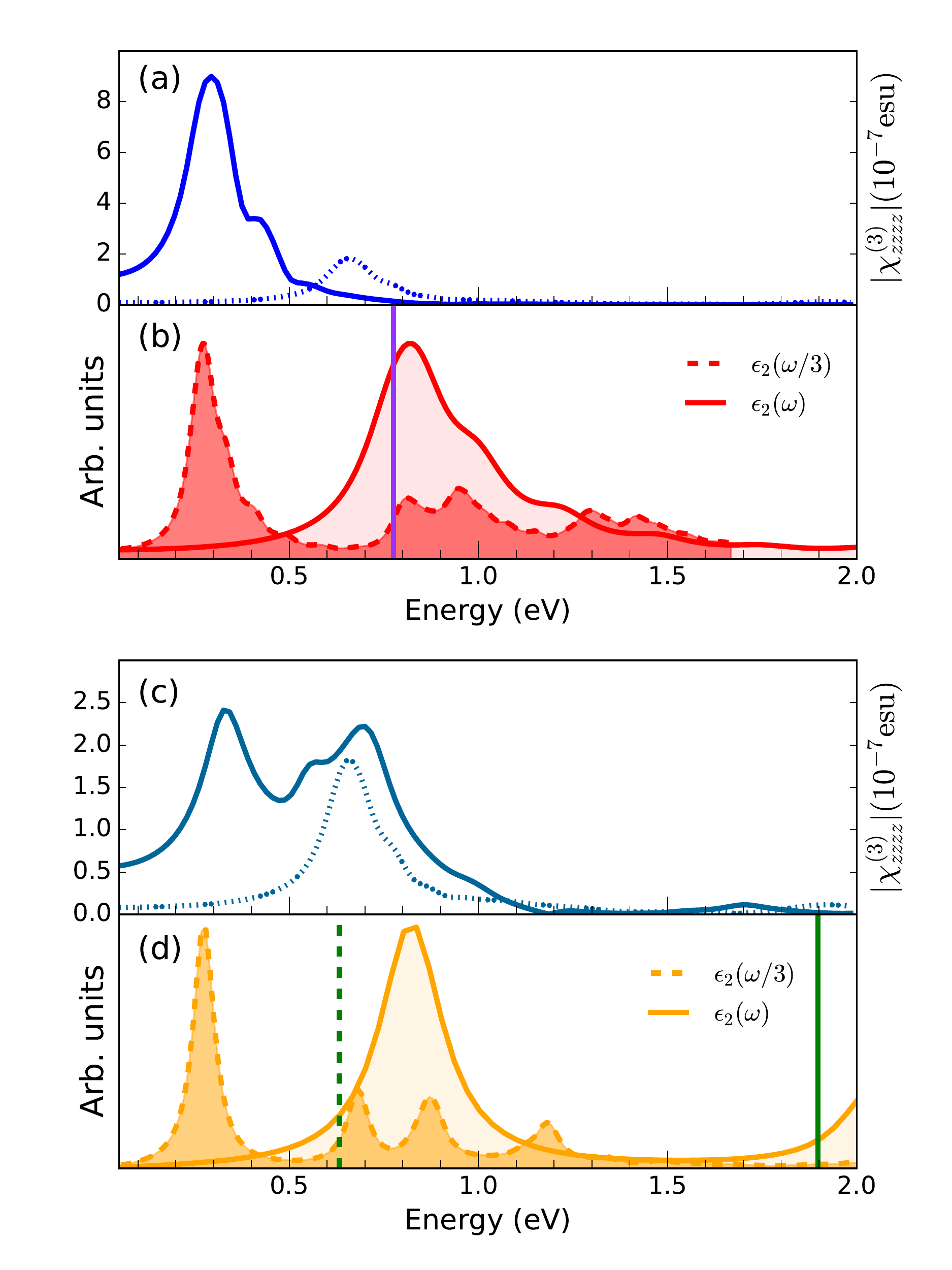}
\caption{[Color online] THG intensity $|\chi^{(3)}_{zzzz}(\omega)|$ and optical absorption in the longitudinal direction ($z$-axis) for 9-AGNR. Panel (a): THG from the IP model (continuous line) and the QP model (dotted line) for the Hamiltonian; panel (b): the imaginary part of the dielectric function at $\omega$ (continuous line) and $\omega/3$ (dashed line) from the IP model for the Hamiltonian; panel (c): THG within the $GW$+BSE model (continuous line) and QP model (dotted line) Hamiltonian;  panel (d): the imaginary part of the dielectric function at $\omega$ (continuous line) and  $\omega/3$ (dashed line) at the same level of approximation. The vertical greeen (violet) lines represent the GW (Kohn-Sham) fundamental gap (continuous line) and the green dashed line represents $1/3$ of the $GW$ fundamental gap.\label{figANGR9}}
\end{figure}

We analyze the many-body effects on the THG [panels (a) and (c) of Fig.\ref{figANGR9}] and on the absorption spectrum [panels (b) and  (d) of the same figure] for the armchair GNR with $N = 9$ (9-AGNR).
Within the IP level of approximation [Eq.\eqref{rt-ip}] the optical spectrum shows the characteristic 1D van Hove singularity at about $0.8$~eV [panels (b) of Fig.\ref{figANGR9}] resulting from transitions at the $\Gamma$ point [see band structure,  panel (c) of Fig.~\ref{fig:bands}] and two shoulders at about $1.0$~eV and $1.2$~eV. 
When the quasiparticle corrections and the electron-hole interaction are turned on [$GW$+BSE model in Eq.~\eqref{rt-bse}] the absorption spectrum below $2$~eV [panels (d) of Fig.\ref{figANGR9}] presents a single excitonic peak well below the onset of the continuum [which is shifted of $1.2$~eV  by quasiparticle corrections, see vertical lines in panels (b) and (d)] with an exceedingly large excitonic binding energy for semiconducting materials ($1$~eV). The main peak is enhanced by about 40\% by the electron-hole interaction.\cite{PhysRevB.84.041401}  To sum up, the inclusion of the quasiparticle corrections and electron-hole interaction modifies the absorption line-shape narrowing the main peak. These effects are indeed known to be important for a qualitative and quantitative predictions of the optical spectra.\cite{PhysRevB.77.041404}

The effect of correlation on the THG intensity is even more dramatic and not entirely predictable from what observed for absorption. 
Panel (a) of Fig.~\ref{figANGR9} shows the third-harmonic intensity obtained from the IP-model. The peaks observed  in the $\chi^{(3)}_{zzzz}$ correspond to three photon-resonances of the van-Hove singularities observed in the absorption and in fact the spectral shape roughly resembles that of the absorption [panel (b) of Fig~\ref{figANGR9}]. When we apply QP correction  to the KS band structure [QP model in Eq.~\eqref{rt-qp}], the spectrum (same panel) is shifted to higher energies by about $0.35$~eV which is roughly 1/3 of the quasiparticle correction to the fundamental band gap (see also Ref.~\onlinecite{PhysRevB.84.041401}). Since by construction $\chi^{(3)}$ contains the product of terms with poles at different energies the $GW$ correction effect is not a simple energy shift. The spectral shape is modified too. More strikingly there is a substantial reduction of the peak intensity (about 75\%) as one can expect from sum rule constraints.\cite{saarinen2002sum}

When excitonic effects are turned on ($GW$+BSE model, panel (c) of Fig~\ref{figANGR9}), the spectra is red-shifted with respect to the QP model. As observed for the absorption, the cancellation between the QP corrections and the exciton binding energy is partial and the main peak in the spectrum is slightly blue-shifted ($0.05$~eV) with respect to the IP model. More importantly the spectral weight is redistributed among few excitonic three-photon resonances and the spectral shape is distinctly different from the one obtained from the IP model. The main peak is in this case significantly broadened.
%The later ones were essentially invisible at the IPA levels. The overall result is different from what we and other groups found for example in second harmonic generation of low dimensional system.\cite{gruning2014}  In fact in the SHG case excitonic effect strongly enhance the nonlinear response both at $\omega$ and $2\omega$. Instead here the final outcome is a redistribution of the spectra at higher frequency with consequent suppression of the response at lower frequencies.(panel d) of Fig.~\ref{figANGR9}).

In analogy to the CNTs, we study how the THG varies as a function of the ribbon size (width) for $N = 3p + 1$, with $p=2,3,4$ [Fig.~\ref{fig:gnrs} panel (b)] and $N = 3p$ with $p=3,4,5$ [Fig.~\ref{fig:gnrs} panel (c)].
Similarly to what is found for the CNTs the intensity increases with the size.  This is a consequence of the bandgap decreasing when the width increases consistently with the findings in Ref.~\onlinecite{ahmadia2011theoretical}. We also observe that the THG of the $N = 3p + 1$ family, which has the largest bandgaps, is smaller by about one-two orders of magnitude than the THG of the $N = 3p$ family, which is related to the suppresion of the intrabands terms in the THG.
As well in the $N = 3p + 1$ the increase of the THG with the size is more pronounced since the bandgap decreases rapidly with $p$. For both families the increase is not uniform, but is larger for the first peak. In particular for the $N =3p$ family the first peak increases by a factor $2$ passing from $N = 9$ to $N =12$ and by a factor $1.5$ passing from $N = 12$ to $N =15$. In both cases it is significantly blue-shifted. The shoulder at $0.4$~eV in the 9-AGNR instead is only slightly blue-shifted and increased in intensity in the larger GNRs.

%\MG{Is there a way to link this to the electronic structure of the GNRs?} These results demonstrate once again that for a detailed understanding of the THG in these carbon nanostructure a simple two-band model may not be sufficient.     
% No idea for the momemnt

Finally our calculations predict a THG of the order of $10^{-7} esu$ at resonances for small GNRs of the same order of magnitude as the THG in graphene.\cite{kumar2013third,hong2013optical,SLGTHG}

%response decreses with increasing gap as it was found in 
%In Fig.~\ref{} we report the THG response for two of the GNR families as function of the ribbon size. The main resonant peak of the THG moves close to zero frequency as the gap decreases and increases its intensity. In order to investigate how correlation effects modify this independent particle picture we investigate the $N=9$ armchair nanoribbon  within our real-time $GW$+BSE approach. The $N=9$ nanoribbon (9-ANGR) belongs to the intermediate case between the nanoribbons with a large and small gaps. TO BE REWRITTEN

\section{Discussion and Conclusions}
\label{conclusions}                                        
Using state-of-the-art ab-initio MBPT approaches ($GW$+BSE) we have studied the THG of two paradigmatic one-dimensional semiconducting nanostructures, a carbon nanotube and a graphene nanoribbon. By comparing the results from the QP and $GW$+BSE model Hamiltonians with the simple IP model Hamiltonian we were able to single out the effect of the electron-electron and electron-hole interaction in the THG of these two systems. For both systems the inclusion of many-body effects modifies significantly the THG: first, the intensity of the main peak is reduced by about a factor 3; second, additional structures have a significant spectral weight and as a result the spectrum has a much richer structure and covers a much larger range of frequencies than its counterpart at the IP level. These results indicate that it is important to include many-body effects for a qualitative and quantitative description of nonlinear optical properties of nanostructures.

Our results agree with what observed by Lacivita and coworkers in a recent study on static hyperpolarizabilities of carbon nanotubes using a coupled perturbed KS approach.~\cite{lacivita2016longitudinal} Their calculations show a gradual suppression of the longitudinal hyperpolarizabilities of the carbon nanotube when increasing the non-local exchange contribution (Hartree-Fock) to the exchange-correlation potential from $0$ to $100$\%. In fact the inclusion of non-local exchange in the DFT functional opens the fundamental band gap---similarly to the QP corrections in our work---which reduces the third-harmonic intensity. Lacivita and coworkers further observed that the reduction due to the band gap opening is only partially compensated by the inclusion of electron-hole interaction in the response function, in agreement with our findings.
Note that in experiments it is difficult, if not impossible to isolate the effects contributing to the THG. For example one could in principle compare the $\chi^{(3)}_{zzzz}$ measured in an isolated CNT with that in CNTs bundles---increasing the dielectric screening and thus `suppressing' the many-body effects. However other effects, as for instance the variation of the phase relaxation time with the environment, would make the interpretation of these experiments not straighforward.\cite{PhysRevLett.94.047404}

Remarkably, including QP corrections and electron-hole interaction has a very different effect on nanostructures optical absorption spectra: QP corrections usually correspond to a rigid blue-shift of the spectra, while electron-hole interaction produces a red-shift and a substantial enhancement of the intensity of the first peak, which is usually more intense than the corresponding van Hove singularity at the IP level. Calculations of the second-harmonic generation\cite{attaccalite2015strong,pedersen2007optical} showed that electron-hole interaction enhances by a 20\% to 200\% the intensity of the main spectral features when compared with spectra calculated at IP level. In the case of SHG in fact addition of QP corrections shifts the spectrum and reduces the overall spectral intensity (as in the case of THG), however this reduction is overcompensated by the enhancement from the electron-hole interaction. 

We argue that the different behavior of the THG, SHG and absorption follows from sum rule constraints.\cite{saarinen2002sum} On the other hand dimensionality could also play an important role since in one-dimensional systems both QP corrections and exciton binding energy are particularly large due to geometrical confinement and poor screening. 
At this regard it would be of interest to systematically study the THG for systems of different dimensionality. At present this study is hindered by the computational cost of solving the time-dependent Bethe-Salpeter equation\cite{attaccalite} for systems with a large number of $\kk$-grid points, as it is the case of bulk semiconductors. 
Other approaches, based for example on the generalization of time-dependent density functional theory\cite{gruning2016dielectrics,PhysRevB.82.235201} or extension of coupled Kohn-Sham equations to dynamical electric fields\cite{newcrystal} could make this kind of studies affordable, once accuracy issues have been addressed. 

Finally, we have also studied the dependence of the THG on the size of the nanostructures (i.e. the radius of the CNT and the width of the GNR). Consistently with previous studies we have found that THG increases by increasing the radius and width of the nanostructures mainly because of the shrinking of the bandgap. Our results show as well that the two-band model that has been used in previous studies does not capture the details of the THG spectrum and more bands close to the Fermi energy need to be taken into account.
%Differently from the linear response case the correction introduced by the $GW$ and the Bethe-Salpeter equation does not compensate each other. The final spectra acquires a broaden shape and loses part of its intensity due to the sum-rules constraints.\cite{saarinen2002sum}

%% This result is quite unexpected in fact in the case of other nonlinear response functions as second-harmonic generation the electron-hole interaction tends to increase the response and not to suppress it.
%% It should be interesting to investigate if this reduction is present only in one-dimensional systems or it is a general trend of the THG response. More studies are necessary to explore this trail.
%% Unfortunately the time-dependent Bethe-Salpeter equation\cite{attaccalite} is still too computational expensive to be applied in bulk materials where a large number of k-points is required, but other strategies
\section{Acknowledgments} 
Computing time has been provided by the national GENGI-IDRIS supercomputing centers at Orsay under contract $n^o$ i2012096655. CA and MG thank the EUspec COST Action for Short Term Scientific Missions that allowed to initiate and carry out this project. CA acknowledges support from the ``Fondation Aix-Marseille Universit\'e".
\addcontentsline{toc}{chapter}{Bibliography}
%\bibliographystyle{apsrev4-1}
%\bibliography{tube,corrnlinear2d,../Theory/nloptics}

\begin{thebibliography}{54}%
\makeatletter
\providecommand \@ifxundefined [1]{%
 \@ifx{#1\undefined}
}%
\providecommand \@ifnum [1]{%
 \ifnum #1\expandafter \@firstoftwo
 \else \expandafter \@secondoftwo
 \fi
}%
\providecommand \@ifx [1]{%
 \ifx #1\expandafter \@firstoftwo
 \else \expandafter \@secondoftwo
 \fi
}%
\providecommand \natexlab [1]{#1}%
\providecommand \enquote  [1]{``#1''}%
\providecommand \bibnamefont  [1]{#1}%
\providecommand \bibfnamefont [1]{#1}%
\providecommand \citenamefont [1]{#1}%
\providecommand \href@noop [0]{\@secondoftwo}%
\providecommand \href [0]{\begingroup \@sanitize@url \@href}%
\providecommand \@href[1]{\@@startlink{#1}\@@href}%
\providecommand \@@href[1]{\endgroup#1\@@endlink}%
\providecommand \@sanitize@url [0]{\catcode `\\12\catcode `\$12\catcode
  `\&12\catcode `\#12\catcode `\^12\catcode `\_12\catcode `\%12\relax}%
\providecommand \@@startlink[1]{}%
\providecommand \@@endlink[0]{}%
\providecommand \url  [0]{\begingroup\@sanitize@url \@url }%
\providecommand \@url [1]{\endgroup\@href {#1}{\urlprefix }}%
\providecommand \urlprefix  [0]{URL }%
\providecommand \Eprint [0]{\href }%
\providecommand \doibase [0]{http://dx.doi.org/}%
\providecommand \selectlanguage [0]{\@gobble}%
\providecommand \bibinfo  [0]{\@secondoftwo}%
\providecommand \bibfield  [0]{\@secondoftwo}%
\providecommand \translation [1]{[#1]}%
\providecommand \BibitemOpen [0]{}%
\providecommand \bibitemStop [0]{}%
\providecommand \bibitemNoStop [0]{.\EOS\space}%
\providecommand \EOS [0]{\spacefactor3000\relax}%
\providecommand \BibitemShut  [1]{\csname bibitem#1\endcsname}%
\let\auto@bib@innerbib\@empty
%</preamble>
\bibitem [{\citenamefont {Wang}\ \emph {et~al.}(2005)\citenamefont {Wang},
  \citenamefont {Dukovic}, \citenamefont {Brus},\ and\ \citenamefont
  {Heinz}}]{Wang838}%
  \BibitemOpen
  \bibfield  {author} {\bibinfo {author} {\bibfnamefont {Feng}\ \bibnamefont
  {Wang}}, \bibinfo {author} {\bibfnamefont {Gordana}\ \bibnamefont {Dukovic}},
  \bibinfo {author} {\bibfnamefont {Louis~E.}\ \bibnamefont {Brus}}, \ and\
  \bibinfo {author} {\bibfnamefont {Tony~F.}\ \bibnamefont {Heinz}},\
  }\bibfield  {title} {\enquote {\bibinfo {title} {The optical resonances in
  carbon nanotubes arise from excitons},}\ }\href {\doibase
  10.1126/science.1110265} {\bibfield  {journal} {\bibinfo  {journal}
  {Science}\ }\textbf {\bibinfo {volume} {308}},\ \bibinfo {pages} {838--841}
  (\bibinfo {year} {2005})}\BibitemShut {NoStop}%
\bibitem [{\citenamefont {Denk}\ \emph {et~al.}(2014)\citenamefont {Denk},
  \citenamefont {Hohage}, \citenamefont {Zeppenfeld}, \citenamefont {Cai},
  \citenamefont {Pignedoli}, \citenamefont {S{\"o}de}, \citenamefont {Fasel},
  \citenamefont {Feng}, \citenamefont {M{\"u}llen}, \citenamefont {Wang} \emph
  {et~al.}}]{denk2014exciton}%
  \BibitemOpen
  \bibfield  {author} {\bibinfo {author} {\bibfnamefont {Richard}\ \bibnamefont
  {Denk}}, \bibinfo {author} {\bibfnamefont {Michael}\ \bibnamefont {Hohage}},
  \bibinfo {author} {\bibfnamefont {Peter}\ \bibnamefont {Zeppenfeld}},
  \bibinfo {author} {\bibfnamefont {Jinming}\ \bibnamefont {Cai}}, \bibinfo
  {author} {\bibfnamefont {Carlo~A}\ \bibnamefont {Pignedoli}}, \bibinfo
  {author} {\bibfnamefont {Hajo}\ \bibnamefont {S{\"o}de}}, \bibinfo {author}
  {\bibfnamefont {Roman}\ \bibnamefont {Fasel}}, \bibinfo {author}
  {\bibfnamefont {Xinliang}\ \bibnamefont {Feng}}, \bibinfo {author}
  {\bibfnamefont {Klaus}\ \bibnamefont {M{\"u}llen}}, \bibinfo {author}
  {\bibfnamefont {Shudong}\ \bibnamefont {Wang}},  \emph {et~al.},\ }\bibfield
  {title} {\enquote {\bibinfo {title} {Exciton-dominated optical response of
  ultra-narrow graphene nanoribbons},}\ }\href@noop {} {\bibfield  {journal}
  {\bibinfo  {journal} {Nature Comm.}\ }\textbf {\bibinfo {volume} {5}}
  (\bibinfo {year} {2014})}\BibitemShut {NoStop}%
\bibitem [{\citenamefont {Prezzi}\ \emph {et~al.}(2008)\citenamefont {Prezzi},
  \citenamefont {Varsano}, \citenamefont {Ruini}, \citenamefont {Marini},\ and\
  \citenamefont {Molinari}}]{PhysRevB.77.041404}%
  \BibitemOpen
  \bibfield  {author} {\bibinfo {author} {\bibfnamefont {Deborah}\ \bibnamefont
  {Prezzi}}, \bibinfo {author} {\bibfnamefont {Daniele}\ \bibnamefont
  {Varsano}}, \bibinfo {author} {\bibfnamefont {Alice}\ \bibnamefont {Ruini}},
  \bibinfo {author} {\bibfnamefont {Andrea}\ \bibnamefont {Marini}}, \ and\
  \bibinfo {author} {\bibfnamefont {Elisa}\ \bibnamefont {Molinari}},\
  }\bibfield  {title} {\enquote {\bibinfo {title} {Optical properties of
  graphene nanoribbons: The role of many-body effects},}\ }\href {\doibase
  10.1103/PhysRevB.77.041404} {\bibfield  {journal} {\bibinfo  {journal} {Phys.
  Rev. B}\ }\textbf {\bibinfo {volume} {77}},\ \bibinfo {pages} {041404}
  (\bibinfo {year} {2008})}\BibitemShut {NoStop}%
\bibitem [{\citenamefont {Spataru}\ \emph {et~al.}(2004)\citenamefont
  {Spataru}, \citenamefont {Ismail-Beigi}, \citenamefont {Benedict},\ and\
  \citenamefont {Louie}}]{PhysRevLett.92.077402}%
  \BibitemOpen
  \bibfield  {author} {\bibinfo {author} {\bibfnamefont {Catalin~D.}\
  \bibnamefont {Spataru}}, \bibinfo {author} {\bibfnamefont {Sohrab}\
  \bibnamefont {Ismail-Beigi}}, \bibinfo {author} {\bibfnamefont {Lorin~X.}\
  \bibnamefont {Benedict}}, \ and\ \bibinfo {author} {\bibfnamefont
  {Steven~G.}\ \bibnamefont {Louie}},\ }\bibfield  {title} {\enquote {\bibinfo
  {title} {Excitonic effects and optical spectra of single-walled carbon
  nanotubes},}\ }\href@noop {} {\bibfield  {journal} {\bibinfo  {journal}
  {Phys. Rev. Lett.}\ }\textbf {\bibinfo {volume} {92}},\ \bibinfo {pages}
  {077402} (\bibinfo {year} {2004})}\BibitemShut {NoStop}%
\bibitem [{\citenamefont {Strinati}(1988)}]{strinati}%
  \BibitemOpen
  \bibfield  {author} {\bibinfo {author} {\bibfnamefont {G.}~\bibnamefont
  {Strinati}},\ }\href@noop {} {\bibfield  {journal} {\bibinfo  {journal} {Riv.
  Nuovo Cimento}\ }\textbf {\bibinfo {volume} {11}},\ \bibinfo {pages} {1--86}
  (\bibinfo {year} {1988})}\BibitemShut {NoStop}%
\bibitem [{\citenamefont {Aulbur}\ \emph {et~al.}(1999)\citenamefont {Aulbur},
  \citenamefont {J{\"o}nsson},\ and\ \citenamefont
  {Wilkins}}]{aulbur1999quasiparticle}%
  \BibitemOpen
  \bibfield  {author} {\bibinfo {author} {\bibfnamefont {Wilfried~G}\
  \bibnamefont {Aulbur}}, \bibinfo {author} {\bibfnamefont {Lars}\ \bibnamefont
  {J{\"o}nsson}}, \ and\ \bibinfo {author} {\bibfnamefont {John~W}\
  \bibnamefont {Wilkins}},\ }\bibfield  {title} {\enquote {\bibinfo {title}
  {Quasiparticle calculations in solids},}\ }\href@noop {} {\bibfield
  {journal} {\bibinfo  {journal} {Solid State Phys.}\ }\textbf {\bibinfo
  {volume} {54}},\ \bibinfo {pages} {1--218} (\bibinfo {year}
  {1999})}\BibitemShut {NoStop}%
\bibitem [{\citenamefont {Chen}\ \emph {et~al.}(2002)\citenamefont {Chen},
  \citenamefont {Raravikar}, \citenamefont {Schadler}, \citenamefont {Ajayan},
  \citenamefont {Zhao}, \citenamefont {Lu}, \citenamefont {Wang},\ and\
  \citenamefont {Zhang}}]{chen2002ultrafast}%
  \BibitemOpen
  \bibfield  {author} {\bibinfo {author} {\bibfnamefont {Y-C}\ \bibnamefont
  {Chen}}, \bibinfo {author} {\bibfnamefont {NR}~\bibnamefont {Raravikar}},
  \bibinfo {author} {\bibfnamefont {LS}~\bibnamefont {Schadler}}, \bibinfo
  {author} {\bibfnamefont {PM}~\bibnamefont {Ajayan}}, \bibinfo {author}
  {\bibfnamefont {Y-P}\ \bibnamefont {Zhao}}, \bibinfo {author} {\bibfnamefont
  {T-M}\ \bibnamefont {Lu}}, \bibinfo {author} {\bibfnamefont {G-C}\
  \bibnamefont {Wang}}, \ and\ \bibinfo {author} {\bibfnamefont {X-C}\
  \bibnamefont {Zhang}},\ }\bibfield  {title} {\enquote {\bibinfo {title}
  {Ultrafast optical switching properties of single-wall carbon nanotube
  polymer composites at 1.55 $\mu$m},}\ }\href@noop {} {\bibfield  {journal}
  {\bibinfo  {journal} {Appl. Phys Lett.}\ }\textbf {\bibinfo {volume} {81}},\
  \bibinfo {pages} {975--977} (\bibinfo {year} {2002})}\BibitemShut {NoStop}%
\bibitem [{\citenamefont {Tatsuura}\ \emph {et~al.}(2003)\citenamefont
  {Tatsuura}, \citenamefont {Furuki}, \citenamefont {Sato}, \citenamefont
  {Iwasa}, \citenamefont {Tian},\ and\ \citenamefont
  {Mitsu}}]{tatsuura2003semiconductor}%
  \BibitemOpen
  \bibfield  {author} {\bibinfo {author} {\bibfnamefont {Satoshi}\ \bibnamefont
  {Tatsuura}}, \bibinfo {author} {\bibfnamefont {Makoto}\ \bibnamefont
  {Furuki}}, \bibinfo {author} {\bibfnamefont {Yasuhiro}\ \bibnamefont {Sato}},
  \bibinfo {author} {\bibfnamefont {Izumi}\ \bibnamefont {Iwasa}}, \bibinfo
  {author} {\bibfnamefont {MINQUAN}\ \bibnamefont {Tian}}, \ and\ \bibinfo
  {author} {\bibfnamefont {Hiroyuki}\ \bibnamefont {Mitsu}},\ }\bibfield
  {title} {\enquote {\bibinfo {title} {Semiconductor carbon nanotubes as
  ultrafast switching materials for optical telecommunications},}\ }\href@noop
  {} {\bibfield  {journal} {\bibinfo  {journal} {Adv. Mat.}\ }\textbf {\bibinfo
  {volume} {15}},\ \bibinfo {pages} {534--537} (\bibinfo {year}
  {2003})}\BibitemShut {NoStop}%
\bibitem [{Note1()}]{Note1}%
  \BibitemOpen
  \bibinfo {note} {See for example Y. R. Shen ``The Principles of Nonlinear
  Optics'', Wiley Interscience (2003)}\BibitemShut {NoStop}%
\bibitem [{\citenamefont {Maeda}\ \emph
  {et~al.}(2005{\natexlab{a}})\citenamefont {Maeda}, \citenamefont {Matsumoto},
  \citenamefont {Kishida}, \citenamefont {Takenobu}, \citenamefont {Iwasa},
  \citenamefont {Shiraishi}, \citenamefont {Ata},\ and\ \citenamefont
  {Okamoto}}]{maeda2005large}%
  \BibitemOpen
  \bibfield  {author} {\bibinfo {author} {\bibfnamefont {Atsushi}\ \bibnamefont
  {Maeda}}, \bibinfo {author} {\bibfnamefont {Shinji}\ \bibnamefont
  {Matsumoto}}, \bibinfo {author} {\bibfnamefont {Hideo}\ \bibnamefont
  {Kishida}}, \bibinfo {author} {\bibfnamefont {Taishi}\ \bibnamefont
  {Takenobu}}, \bibinfo {author} {\bibfnamefont {Yoshihiro}\ \bibnamefont
  {Iwasa}}, \bibinfo {author} {\bibfnamefont {M}~\bibnamefont {Shiraishi}},
  \bibinfo {author} {\bibfnamefont {M}~\bibnamefont {Ata}}, \ and\ \bibinfo
  {author} {\bibfnamefont {H}~\bibnamefont {Okamoto}},\ }\bibfield  {title}
  {\enquote {\bibinfo {title} {Large optical nonlinearity of semiconducting
  single-walled carbon nanotubes under resonant excitations},}\ }\href@noop {}
  {\bibfield  {journal} {\bibinfo  {journal} {Physical review letters}\
  }\textbf {\bibinfo {volume} {94}},\ \bibinfo {pages} {047404} (\bibinfo
  {year} {2005}{\natexlab{a}})}\BibitemShut {NoStop}%
\bibitem [{\citenamefont {Liu}\ \emph {et~al.}(1999)\citenamefont {Liu},
  \citenamefont {Si}, \citenamefont {Chang}, \citenamefont {Xu}, \citenamefont
  {Yang}, \citenamefont {Pan}, \citenamefont {Xie}, \citenamefont {Ye},
  \citenamefont {Fan},\ and\ \citenamefont {Wan}}]{liu1999third}%
  \BibitemOpen
  \bibfield  {author} {\bibinfo {author} {\bibfnamefont {Xuchun}\ \bibnamefont
  {Liu}}, \bibinfo {author} {\bibfnamefont {Jinhai}\ \bibnamefont {Si}},
  \bibinfo {author} {\bibfnamefont {Baohe}\ \bibnamefont {Chang}}, \bibinfo
  {author} {\bibfnamefont {Gang}\ \bibnamefont {Xu}}, \bibinfo {author}
  {\bibfnamefont {Qiguang}\ \bibnamefont {Yang}}, \bibinfo {author}
  {\bibfnamefont {Zhengwei}\ \bibnamefont {Pan}}, \bibinfo {author}
  {\bibfnamefont {Sishen}\ \bibnamefont {Xie}}, \bibinfo {author}
  {\bibfnamefont {Peixian}\ \bibnamefont {Ye}}, \bibinfo {author}
  {\bibfnamefont {Junhua}\ \bibnamefont {Fan}}, \ and\ \bibinfo {author}
  {\bibfnamefont {Meixiang}\ \bibnamefont {Wan}},\ }\bibfield  {title}
  {\enquote {\bibinfo {title} {Third-order optical nonlinearity of the carbon
  nanotubes},}\ }\href@noop {} {\bibfield  {journal} {\bibinfo  {journal}
  {Appl. Phys. Lett.}\ }\textbf {\bibinfo {volume} {74}},\ \bibinfo {pages}
  {164--166} (\bibinfo {year} {1999})}\BibitemShut {NoStop}%
\bibitem [{\citenamefont {Ichida}\ \emph {et~al.}(2009)\citenamefont {Ichida},
  \citenamefont {Kiyohara}, \citenamefont {Saito}, \citenamefont {Miyata},
  \citenamefont {Kataura},\ and\ \citenamefont {Ando}}]{Ichida20091794}%
  \BibitemOpen
  \bibfield  {author} {\bibinfo {author} {\bibfnamefont {Masao}\ \bibnamefont
  {Ichida}}, \bibinfo {author} {\bibfnamefont {Yumie}\ \bibnamefont
  {Kiyohara}}, \bibinfo {author} {\bibfnamefont {Shingo}\ \bibnamefont
  {Saito}}, \bibinfo {author} {\bibfnamefont {Yasumitsu}\ \bibnamefont
  {Miyata}}, \bibinfo {author} {\bibfnamefont {Hiromichi}\ \bibnamefont
  {Kataura}}, \ and\ \bibinfo {author} {\bibfnamefont {Hiroaki}\ \bibnamefont
  {Ando}},\ }\bibfield  {title} {\enquote {\bibinfo {title} {Nonlinear optical
  properties and phase-relaxation processes in single-walled carbon
  nanotubes},}\ }\href {\doibase
  http://dx.doi.org/10.1016/j.jlumin.2009.04.066} {\bibfield  {journal}
  {\bibinfo  {journal} {Journal of Luminescence}\ }\textbf {\bibinfo {volume}
  {129}},\ \bibinfo {pages} {1794 -- 1797} (\bibinfo {year} {2009})},\ \bibinfo
  {note} {special Issue based on The 15th International Conference on
  Luminescence and Optical Spectroscopy of Condensed Matter
  (ICL'08)}\BibitemShut {NoStop}%
\bibitem [{\citenamefont {Lauret}\ \emph {et~al.}(2004)\citenamefont {Lauret},
  \citenamefont {Voisin}, \citenamefont {Cassabois}, \citenamefont {Tignon},
  \citenamefont {Delalande}, \citenamefont {Roussignol}, \citenamefont {Jost},\
  and\ \citenamefont {Capes}}]{lauret2004third}%
  \BibitemOpen
  \bibfield  {author} {\bibinfo {author} {\bibfnamefont {Jean-S{\'e}bastien}\
  \bibnamefont {Lauret}}, \bibinfo {author} {\bibfnamefont {Christophe}\
  \bibnamefont {Voisin}}, \bibinfo {author} {\bibfnamefont {Guillaume}\
  \bibnamefont {Cassabois}}, \bibinfo {author} {\bibfnamefont {J{\'e}r{\^o}me}\
  \bibnamefont {Tignon}}, \bibinfo {author} {\bibfnamefont {Claude}\
  \bibnamefont {Delalande}}, \bibinfo {author} {\bibfnamefont {Ph}~\bibnamefont
  {Roussignol}}, \bibinfo {author} {\bibfnamefont {Oliver}\ \bibnamefont
  {Jost}}, \ and\ \bibinfo {author} {\bibfnamefont {L}~\bibnamefont {Capes}},\
  }\bibfield  {title} {\enquote {\bibinfo {title} {Third-order optical
  nonlinearities of carbon nanotubes in the femtosecond regime},}\ }\href@noop
  {} {\bibfield  {journal} {\bibinfo  {journal} {Appl. Phys. Lett.}\ }\textbf
  {\bibinfo {volume} {85}},\ \bibinfo {pages} {3572--3574} (\bibinfo {year}
  {2004})}\BibitemShut {NoStop}%
\bibitem [{\citenamefont {Stanciu}\ \emph {et~al.}(2002)\citenamefont
  {Stanciu}, \citenamefont {Ehlich}, \citenamefont {Petrov}, \citenamefont
  {Steinkellner}, \citenamefont {Herrmann}, \citenamefont {Hertel},
  \citenamefont {Slepyan}, \citenamefont {Khrutchinski}, \citenamefont
  {Maksimenko}, \citenamefont {Rotermund} \emph
  {et~al.}}]{stanciu2002experimental}%
  \BibitemOpen
  \bibfield  {author} {\bibinfo {author} {\bibfnamefont {C}~\bibnamefont
  {Stanciu}}, \bibinfo {author} {\bibfnamefont {R}~\bibnamefont {Ehlich}},
  \bibinfo {author} {\bibfnamefont {V}~\bibnamefont {Petrov}}, \bibinfo
  {author} {\bibfnamefont {O}~\bibnamefont {Steinkellner}}, \bibinfo {author}
  {\bibfnamefont {J}~\bibnamefont {Herrmann}}, \bibinfo {author} {\bibfnamefont
  {IV}~\bibnamefont {Hertel}}, \bibinfo {author} {\bibfnamefont {G~Ya}\
  \bibnamefont {Slepyan}}, \bibinfo {author} {\bibfnamefont {AA}~\bibnamefont
  {Khrutchinski}}, \bibinfo {author} {\bibfnamefont {SA}~\bibnamefont
  {Maksimenko}}, \bibinfo {author} {\bibfnamefont {F}~\bibnamefont
  {Rotermund}},  \emph {et~al.},\ }\bibfield  {title} {\enquote {\bibinfo
  {title} {Experimental and theoretical study of third-order harmonic
  generation in carbon nanotubes},}\ }\href@noop {} {\bibfield  {journal}
  {\bibinfo  {journal} {Appl. Phys. lett.}\ }\textbf {\bibinfo {volume} {81}},\
  \bibinfo {pages} {4064--4066} (\bibinfo {year} {2002})}\BibitemShut {NoStop}%
\bibitem [{\citenamefont {Nemilentsau}\ \emph
  {et~al.}(2006{\natexlab{a}})\citenamefont {Nemilentsau}, \citenamefont
  {Slepyan}, \citenamefont {Khrutchinskii},\ and\ \citenamefont
  {Maksimenko}}]{nemilentsau2006third}%
  \BibitemOpen
  \bibfield  {author} {\bibinfo {author} {\bibfnamefont {AM}~\bibnamefont
  {Nemilentsau}}, \bibinfo {author} {\bibfnamefont {G~Ya}\ \bibnamefont
  {Slepyan}}, \bibinfo {author} {\bibfnamefont {AA}~\bibnamefont
  {Khrutchinskii}}, \ and\ \bibinfo {author} {\bibfnamefont {SA}~\bibnamefont
  {Maksimenko}},\ }\bibfield  {title} {\enquote {\bibinfo {title} {Third-order
  optical nonlinearity in single-wall carbon nanotubes},}\ }\href@noop {}
  {\bibfield  {journal} {\bibinfo  {journal} {Carbon}\ }\textbf {\bibinfo
  {volume} {44}},\ \bibinfo {pages} {2246--2253} (\bibinfo {year}
  {2006}{\natexlab{a}})}\BibitemShut {NoStop}%
\bibitem [{\citenamefont {Kumar}\ \emph {et~al.}(2013)\citenamefont {Kumar},
  \citenamefont {Kumar}, \citenamefont {Gerstenkorn}, \citenamefont {Wang},
  \citenamefont {Chiu}, \citenamefont {Smirl},\ and\ \citenamefont
  {Zhao}}]{kumar2013third}%
  \BibitemOpen
  \bibfield  {author} {\bibinfo {author} {\bibfnamefont {Nardeep}\ \bibnamefont
  {Kumar}}, \bibinfo {author} {\bibfnamefont {Jatinder}\ \bibnamefont {Kumar}},
  \bibinfo {author} {\bibfnamefont {Chris}\ \bibnamefont {Gerstenkorn}},
  \bibinfo {author} {\bibfnamefont {Rui}\ \bibnamefont {Wang}}, \bibinfo
  {author} {\bibfnamefont {Hsin-Ying}\ \bibnamefont {Chiu}}, \bibinfo {author}
  {\bibfnamefont {Arthur~L}\ \bibnamefont {Smirl}}, \ and\ \bibinfo {author}
  {\bibfnamefont {Hui}\ \bibnamefont {Zhao}},\ }\bibfield  {title} {\enquote
  {\bibinfo {title} {Third harmonic generation in graphene and few-layer
  graphite films},}\ }\href@noop {} {\bibfield  {journal} {\bibinfo  {journal}
  {Physical Review B}\ }\textbf {\bibinfo {volume} {87}},\ \bibinfo {pages}
  {121406} (\bibinfo {year} {2013})}\BibitemShut {NoStop}%
\bibitem [{\citenamefont {Hong}\ \emph {et~al.}(2013)\citenamefont {Hong},
  \citenamefont {Dadap}, \citenamefont {Petrone}, \citenamefont {Yeh},
  \citenamefont {Hone},\ and\ \citenamefont {Osgood~Jr}}]{hong2013optical}%
  \BibitemOpen
  \bibfield  {author} {\bibinfo {author} {\bibfnamefont {Sung-Young}\
  \bibnamefont {Hong}}, \bibinfo {author} {\bibfnamefont {Jerry~I}\
  \bibnamefont {Dadap}}, \bibinfo {author} {\bibfnamefont {Nicholas}\
  \bibnamefont {Petrone}}, \bibinfo {author} {\bibfnamefont {Po-Chun}\
  \bibnamefont {Yeh}}, \bibinfo {author} {\bibfnamefont {James}\ \bibnamefont
  {Hone}}, \ and\ \bibinfo {author} {\bibfnamefont {Richard~M}\ \bibnamefont
  {Osgood~Jr}},\ }\bibfield  {title} {\enquote {\bibinfo {title} {Optical
  third-harmonic generation in graphene},}\ }\href@noop {} {\bibfield
  {journal} {\bibinfo  {journal} {Physical Review X}\ }\textbf {\bibinfo
  {volume} {3}},\ \bibinfo {pages} {021014} (\bibinfo {year}
  {2013})}\BibitemShut {NoStop}%
\bibitem [{\citenamefont {S{\"{a}}yn{\"{a}}tjoki}\ \emph
  {et~al.}(2013)\citenamefont {S{\"{a}}yn{\"{a}}tjoki}, \citenamefont
  {Karvonen}, \citenamefont {Riikonen}, \citenamefont {Kim}, \citenamefont
  {Mehravar}, \citenamefont {Norwood}, \citenamefont {Peyghambarian},
  \citenamefont {Lipsanen},\ and\ \citenamefont {Kieu}}]{SLGTHG}%
  \BibitemOpen
  \bibfield  {author} {\bibinfo {author} {\bibfnamefont {Antti}\ \bibnamefont
  {S{\"{a}}yn{\"{a}}tjoki}}, \bibinfo {author} {\bibfnamefont {Lasse}\
  \bibnamefont {Karvonen}}, \bibinfo {author} {\bibfnamefont {Juha}\
  \bibnamefont {Riikonen}}, \bibinfo {author} {\bibfnamefont {Wonjae}\
  \bibnamefont {Kim}}, \bibinfo {author} {\bibfnamefont {Soroush}\ \bibnamefont
  {Mehravar}}, \bibinfo {author} {\bibfnamefont {Robert~A}\ \bibnamefont
  {Norwood}}, \bibinfo {author} {\bibfnamefont {Nasser}\ \bibnamefont
  {Peyghambarian}}, \bibinfo {author} {\bibfnamefont {Harri}\ \bibnamefont
  {Lipsanen}}, \ and\ \bibinfo {author} {\bibfnamefont {Khanh}\ \bibnamefont
  {Kieu}},\ }\bibfield  {title} {\enquote {\bibinfo {title} {{Rapid Large-Area
  Multiphoton Microscopy for Characterization of Graphene}},}\ }\href@noop {}
  {\bibfield  {journal} {\bibinfo  {journal} {ACS Nano}\ }\textbf {\bibinfo
  {volume} {7}},\ \bibinfo {pages} {8441--8446} (\bibinfo {year}
  {2013})}\BibitemShut {NoStop}%
\bibitem [{\citenamefont {Margulis}\ and\ \citenamefont
  {Sizikova}(1998)}]{margulis1998theoretical}%
  \BibitemOpen
  \bibfield  {author} {\bibinfo {author} {\bibfnamefont {Vl~A}\ \bibnamefont
  {Margulis}}\ and\ \bibinfo {author} {\bibfnamefont {TA}~\bibnamefont
  {Sizikova}},\ }\bibfield  {title} {\enquote {\bibinfo {title} {Theoretical
  study of third-order nonlinear optical response of semiconductor carbon
  nanotubes},}\ }\href@noop {} {\bibfield  {journal} {\bibinfo  {journal}
  {Physica B: Condensed Matter}\ }\textbf {\bibinfo {volume} {245}},\ \bibinfo
  {pages} {173--189} (\bibinfo {year} {1998})}\BibitemShut {NoStop}%
\bibitem [{\citenamefont {Zhang}\ \emph {et~al.}(2006)\citenamefont {Zhang},
  \citenamefont {Guo},\ and\ \citenamefont {Liang}}]{Zhang2006101}%
  \BibitemOpen
  \bibfield  {author} {\bibinfo {author} {\bibfnamefont {Chao-Jin}\
  \bibnamefont {Zhang}}, \bibinfo {author} {\bibfnamefont {Kang-Xian}\
  \bibnamefont {Guo}}, \ and\ \bibinfo {author} {\bibfnamefont {Shi-Dong}\
  \bibnamefont {Liang}},\ }\bibfield  {title} {\enquote {\bibinfo {title}
  {Third harmonic generation of semiconductor carbon nanotubes},}\ }\href
  {\doibase http://dx.doi.org/10.1016/j.cplett.2006.10.125} {\bibfield
  {journal} {\bibinfo  {journal} {Chem. Phys. Lett.}\ }\textbf {\bibinfo
  {volume} {433}},\ \bibinfo {pages} {101 -- 104} (\bibinfo {year}
  {2006})}\BibitemShut {NoStop}%
\bibitem [{\citenamefont {Rezania}\ and\ \citenamefont
  {Daneshfar}(2012)}]{rezania}%
  \BibitemOpen
  \bibfield  {author} {\bibinfo {author} {\bibfnamefont {Hamed}\ \bibnamefont
  {Rezania}}\ and\ \bibinfo {author} {\bibfnamefont {Nader}\ \bibnamefont
  {Daneshfar}},\ }\bibfield  {title} {\enquote {\bibinfo {title} {Study of
  third harmonic generation in zigzag carbon nanotubes using the green's
  function approach},}\ }\href {\doibase 10.1007/s00339-012-7063-7} {\bibfield
  {journal} {\bibinfo  {journal} {Appl. Phys. A}\ }\textbf {\bibinfo {volume}
  {109}},\ \bibinfo {pages} {503--508} (\bibinfo {year} {2012})}\BibitemShut
  {NoStop}%
\bibitem [{\citenamefont {Lacivita}\ \emph {et~al.}(2016)\citenamefont
  {Lacivita}, \citenamefont {R{\'e}rat}, \citenamefont {Orlando}, \citenamefont
  {Dovesi},\ and\ \citenamefont {D’arco}}]{lacivita2016longitudinal}%
  \BibitemOpen
  \bibfield  {author} {\bibinfo {author} {\bibfnamefont {Valentina}\
  \bibnamefont {Lacivita}}, \bibinfo {author} {\bibfnamefont {Michel}\
  \bibnamefont {R{\'e}rat}}, \bibinfo {author} {\bibfnamefont {Roberto}\
  \bibnamefont {Orlando}}, \bibinfo {author} {\bibfnamefont {Roberto}\
  \bibnamefont {Dovesi}}, \ and\ \bibinfo {author} {\bibfnamefont {Philippe}\
  \bibnamefont {D’arco}},\ }\bibfield  {title} {\enquote {\bibinfo {title}
  {Longitudinal and transverse hyperpolarizabilities of carbon nanotubes: a
  computational investigation through the coupled-perturbed
  hartree--fock/kohn--sham scheme},}\ }\href@noop {} {\bibfield  {journal}
  {\bibinfo  {journal} {Theor. Chem. Acc.}\ }\textbf {\bibinfo {volume}
  {135}},\ \bibinfo {pages} {81} (\bibinfo {year} {2016})}\BibitemShut
  {NoStop}%
\bibitem [{Note2()}]{Note2}%
  \BibitemOpen
  \bibinfo {note} {Note that none of the mentioned studied included temperature
  effects, except Ref.~\protect \rev@citealp {rezania} which includes
  temperature effects by means of Green's function theory.}\BibitemShut {Stop}%
\bibitem [{\citenamefont {Wang}\ and\ \citenamefont
  {Andersen}(2016)}]{wang2016first}%
  \BibitemOpen
  \bibfield  {author} {\bibinfo {author} {\bibfnamefont {Yichao}\ \bibnamefont
  {Wang}}\ and\ \bibinfo {author} {\bibfnamefont {David~R.}\ \bibnamefont
  {Andersen}},\ }\bibfield  {title} {\enquote {\bibinfo {title}
  {First-principles study of the terahertz third-order nonlinear response of
  metallic armchair graphene nanoribbons},}\ }\href@noop {} {\bibfield
  {journal} {\bibinfo  {journal} {Phys. Rev. B}\ }\textbf {\bibinfo {volume}
  {93}},\ \bibinfo {pages} {235430} (\bibinfo {year} {2016})}\BibitemShut
  {NoStop}%
\bibitem [{\citenamefont {Zhang}\ \emph {et~al.}(2014)\citenamefont {Zhang},
  \citenamefont {Li},\ and\ \citenamefont {Li}}]{C3TC31847H}%
  \BibitemOpen
  \bibfield  {author} {\bibinfo {author} {\bibfnamefont {Minyi}\ \bibnamefont
  {Zhang}}, \bibinfo {author} {\bibfnamefont {Guangshe}\ \bibnamefont {Li}}, \
  and\ \bibinfo {author} {\bibfnamefont {Liping}\ \bibnamefont {Li}},\
  }\bibfield  {title} {\enquote {\bibinfo {title} {Graphene nanoribbons
  generate a strong third-order nonlinear optical response upon intercalating
  hexagonal boron nitride},}\ }\href {\doibase 10.1039/C3TC31847H} {\bibfield
  {journal} {\bibinfo  {journal} {J. Mater. Chem. C}\ }\textbf {\bibinfo
  {volume} {2}},\ \bibinfo {pages} {1482--1488} (\bibinfo {year}
  {2014})}\BibitemShut {NoStop}%
\bibitem [{\citenamefont {Karamanis}\ and\ \citenamefont
  {Pouchan}(2013)}]{karamanis2013second}%
  \BibitemOpen
  \bibfield  {author} {\bibinfo {author} {\bibfnamefont {Panaghiotis}\
  \bibnamefont {Karamanis}}\ and\ \bibinfo {author} {\bibfnamefont {Claude}\
  \bibnamefont {Pouchan}},\ }\bibfield  {title} {\enquote {\bibinfo {title}
  {Second-hyperpolarizability ($\gamma$) enhancement in metal-decorated zigzag
  graphene flakes and ribbons: The size effect},}\ }\href@noop {} {\bibfield
  {journal} {\bibinfo  {journal} {J. of Phys. Chem. C}\ }\textbf {\bibinfo
  {volume} {117}},\ \bibinfo {pages} {3134--3140} (\bibinfo {year}
  {2013})}\BibitemShut {NoStop}%
\bibitem [{\citenamefont {Ahmadia}\ and\ \citenamefont
  {Asgarib}(2011)}]{ahmadia2011theoretical}%
  \BibitemOpen
  \bibfield  {author} {\bibinfo {author} {\bibfnamefont {E}~\bibnamefont
  {Ahmadia}}\ and\ \bibinfo {author} {\bibfnamefont {A}~\bibnamefont
  {Asgarib}},\ }\href@noop {} {\enquote {\bibinfo {title} {Theoretical
  calculation of third order susceptibility of armchair graphene nanoribbon at
  near infrared range},}\ } (\bibinfo {year} {2011}),\ \bibinfo {note}
  {http://graphita.bo.imm.cnr.it/graphita2011/Poster/Ahmadi.pdf}\BibitemShut
  {NoStop}%
\bibitem [{\citenamefont {Attaccalite}\ \emph {et~al.}(2015)\citenamefont
  {Attaccalite}, \citenamefont {Nguer}, \citenamefont {Cannuccia},\ and\
  \citenamefont {Gr{\"u}ning}}]{attaccalite2015strong}%
  \BibitemOpen
  \bibfield  {author} {\bibinfo {author} {\bibfnamefont {Claudio}\ \bibnamefont
  {Attaccalite}}, \bibinfo {author} {\bibfnamefont {Alassane}\ \bibnamefont
  {Nguer}}, \bibinfo {author} {\bibfnamefont {Elena}\ \bibnamefont
  {Cannuccia}}, \ and\ \bibinfo {author} {\bibfnamefont {Myrta}\ \bibnamefont
  {Gr{\"u}ning}},\ }\bibfield  {title} {\enquote {\bibinfo {title} {Strong
  second harmonic generation in sic, zno, gan two-dimensional hexagonal
  crystals from first-principles many-body calculations},}\ }\href@noop {}
  {\bibfield  {journal} {\bibinfo  {journal} {Phys. Chem. Chem. Phys.}\
  }\textbf {\bibinfo {volume} {17}},\ \bibinfo {pages} {9533--9540} (\bibinfo
  {year} {2015})}\BibitemShut {NoStop}%
\bibitem [{\citenamefont {Gr{\"u}ning}\ and\ \citenamefont
  {Attaccalite}(2014)}]{gruning2014}%
  \BibitemOpen
  \bibfield  {author} {\bibinfo {author} {\bibfnamefont {M}~\bibnamefont
  {Gr{\"u}ning}}\ and\ \bibinfo {author} {\bibfnamefont {Claudio}\ \bibnamefont
  {Attaccalite}},\ }\bibfield  {title} {\enquote {\bibinfo {title} {Second
  harmonic generation in h--bn and mos$_2$ monolayers: Role of electron-hole
  interaction},}\ }\href@noop {} {\bibfield  {journal} {\bibinfo  {journal}
  {Phys. Rev. B}\ }\textbf {\bibinfo {volume} {89}},\ \bibinfo {pages} {081102}
  (\bibinfo {year} {2014})}\BibitemShut {NoStop}%
\bibitem [{Note3()}]{Note3}%
  \BibitemOpen
  \bibinfo {note} {A centrosymmetric system has an inversion center and thus
  the second-harmonic generation is zero within the electric dipole
  approximation. When multipole corrections are considered the second-harmonic
  generation is nonzero, but still very small.}\BibitemShut {Stop}%
\bibitem [{\citenamefont {Perdew}\ and\ \citenamefont
  {Zunger}(1981)}]{PhysRevB.23.5048}%
  \BibitemOpen
  \bibfield  {author} {\bibinfo {author} {\bibfnamefont {J.~P.}\ \bibnamefont
  {Perdew}}\ and\ \bibinfo {author} {\bibfnamefont {Alex}\ \bibnamefont
  {Zunger}},\ }\bibfield  {title} {\enquote {\bibinfo {title} {Self-interaction
  correction to density-functional approximations for many-electron systems},}\
  }\href {\doibase 10.1103/PhysRevB.23.5048} {\bibfield  {journal} {\bibinfo
  {journal} {Phys. Rev. B}\ }\textbf {\bibinfo {volume} {23}},\ \bibinfo
  {pages} {5048--5079} (\bibinfo {year} {1981})}\BibitemShut {NoStop}%
\bibitem [{\citenamefont {Perdew}\ and\ \citenamefont
  {Wang}(1992)}]{PhysRevB.45.13244}%
  \BibitemOpen
  \bibfield  {author} {\bibinfo {author} {\bibfnamefont {John~P.}\ \bibnamefont
  {Perdew}}\ and\ \bibinfo {author} {\bibfnamefont {Yue}\ \bibnamefont
  {Wang}},\ }\bibfield  {title} {\enquote {\bibinfo {title} {Accurate and
  simple analytic representation of the electron-gas correlation energy},}\
  }\href {\doibase 10.1103/PhysRevB.45.13244} {\bibfield  {journal} {\bibinfo
  {journal} {Phys. Rev. B}\ }\textbf {\bibinfo {volume} {45}},\ \bibinfo
  {pages} {13244--13249} (\bibinfo {year} {1992})}\BibitemShut {NoStop}%
\bibitem [{\citenamefont {Giannozzi}\ \emph {et~al.}(2009)\citenamefont
  {Giannozzi} \emph {et~al.}}]{pwscf}%
  \BibitemOpen
  \bibfield  {author} {\bibinfo {author} {\bibfnamefont {P.}~\bibnamefont
  {Giannozzi}} \emph {et~al.},\ }\href@noop {} {\bibfield  {journal} {\bibinfo
  {journal} {J. Phys. Condens. Matter}\ }\textbf {\bibinfo {volume} {21}},\
  \bibinfo {pages} {395502} (\bibinfo {year} {2009})},\ \bibinfo {note}
  {http://www.quantum-espresso.org}\BibitemShut {NoStop}%
\bibitem [{\citenamefont {Troullier}\ and\ \citenamefont
  {Martins}(1991)}]{troullier}%
  \BibitemOpen
  \bibfield  {author} {\bibinfo {author} {\bibfnamefont {N.}~\bibnamefont
  {Troullier}}\ and\ \bibinfo {author} {\bibfnamefont {J.~L.}\ \bibnamefont
  {Martins}},\ }\bibfield  {title} {\enquote {\bibinfo {title} {Efficient
  pseudopotentials for plane-wave calculations},}\ }\href@noop {} {\bibfield
  {journal} {\bibinfo  {journal} {Phys. Rev. B}\ }\textbf {\bibinfo {volume}
  {43}},\ \bibinfo {pages} {1993} (\bibinfo {year} {1991})}\BibitemShut
  {NoStop}%
\bibitem [{\citenamefont {Rojas}\ \emph {et~al.}(1995)\citenamefont {Rojas},
  \citenamefont {Godby},\ and\ \citenamefont {Needs}}]{godby}%
  \BibitemOpen
  \bibfield  {author} {\bibinfo {author} {\bibfnamefont {H.~N.}\ \bibnamefont
  {Rojas}}, \bibinfo {author} {\bibfnamefont {R.~W.}\ \bibnamefont {Godby}}, \
  and\ \bibinfo {author} {\bibfnamefont {R.~J.}\ \bibnamefont {Needs}},\
  }\bibfield  {title} {\enquote {\bibinfo {title} {Space-time method for ab
  initio calculations of self-energies and dielectric response functions of
  solids},}\ }\href {\doibase 10.1103/PhysRevLett.74.1827} {\bibfield
  {journal} {\bibinfo  {journal} {Phys. Rev. Lett.}\ }\textbf {\bibinfo
  {volume} {74}},\ \bibinfo {pages} {1827--1830} (\bibinfo {year}
  {1995})}\BibitemShut {NoStop}%
\bibitem [{\citenamefont {Rozzi}\ \emph {et~al.}(2006)\citenamefont {Rozzi},
  \citenamefont {Varsano}, \citenamefont {Marini}, \citenamefont {Gross},\ and\
  \citenamefont {Rubio}}]{rozzi2006exact}%
  \BibitemOpen
  \bibfield  {author} {\bibinfo {author} {\bibfnamefont {Carlo~A}\ \bibnamefont
  {Rozzi}}, \bibinfo {author} {\bibfnamefont {Daniele}\ \bibnamefont
  {Varsano}}, \bibinfo {author} {\bibfnamefont {Andrea}\ \bibnamefont
  {Marini}}, \bibinfo {author} {\bibfnamefont {Eberhard K.~U.}\ \bibnamefont
  {Gross}}, \ and\ \bibinfo {author} {\bibfnamefont {Angel}\ \bibnamefont
  {Rubio}},\ }\bibfield  {title} {\enquote {\bibinfo {title} {Exact coulomb
  cutoff technique for supercell calculations},}\ }\href@noop {} {\bibfield
  {journal} {\bibinfo  {journal} {Phys. Rev. B}\ }\textbf {\bibinfo {volume}
  {73}},\ \bibinfo {pages} {205119} (\bibinfo {year} {2006})}\BibitemShut
  {NoStop}%
\bibitem [{\citenamefont {Schmidt}\ \emph {et~al.}(2003)\citenamefont
  {Schmidt}, \citenamefont {Glutsch}, \citenamefont {Hahn},\ and\ \citenamefont
  {Bechstedt}}]{schmidt2003efficient}%
  \BibitemOpen
  \bibfield  {author} {\bibinfo {author} {\bibfnamefont {W.~G.}\ \bibnamefont
  {Schmidt}}, \bibinfo {author} {\bibfnamefont {S}~\bibnamefont {Glutsch}},
  \bibinfo {author} {\bibfnamefont {P.~H.}\ \bibnamefont {Hahn}}, \ and\
  \bibinfo {author} {\bibfnamefont {F}~\bibnamefont {Bechstedt}},\ }\bibfield
  {title} {\enquote {\bibinfo {title} {Efficient $o(n^2)$ method to solve the
  bethe-salpeter equation},}\ }\href@noop {} {\bibfield  {journal} {\bibinfo
  {journal} {Phys. Rev. B}\ }\textbf {\bibinfo {volume} {67}},\ \bibinfo
  {pages} {085307} (\bibinfo {year} {2003})}\BibitemShut {NoStop}%
\bibitem [{\citenamefont {Souza}\ \emph {et~al.}(2004)\citenamefont {Souza},
  \citenamefont {\'I\~niguez},\ and\ \citenamefont {Vanderbilt}}]{souza_prb}%
  \BibitemOpen
  \bibfield  {author} {\bibinfo {author} {\bibfnamefont {Ivo}\ \bibnamefont
  {Souza}}, \bibinfo {author} {\bibfnamefont {Jorge}\ \bibnamefont
  {\'I\~niguez}}, \ and\ \bibinfo {author} {\bibfnamefont {David}\ \bibnamefont
  {Vanderbilt}},\ }\bibfield  {title} {\enquote {\bibinfo {title} {Dynamics of
  berry-phase polarization in time-dependent electric fields},}\ }\href@noop {}
  {\bibfield  {journal} {\bibinfo  {journal} {Phys. Rev. B}\ }\textbf {\bibinfo
  {volume} {69}},\ \bibinfo {pages} {085106} (\bibinfo {year}
  {2004})}\BibitemShut {NoStop}%
\bibitem [{\citenamefont {Attaccalite}\ and\ \citenamefont
  {Gr{\"u}ning}(2013)}]{nloptics2013}%
  \BibitemOpen
  \bibfield  {author} {\bibinfo {author} {\bibfnamefont {C}~\bibnamefont
  {Attaccalite}}\ and\ \bibinfo {author} {\bibfnamefont {M}~\bibnamefont
  {Gr{\"u}ning}},\ }\bibfield  {title} {\enquote {\bibinfo {title} {Nonlinear
  optics from an ab initio approach by means of the dynamical berry phase:
  Application to second-and third-harmonic generation in semiconductors},}\
  }\href@noop {} {\bibfield  {journal} {\bibinfo  {journal} {Phys. Rev. B}\
  }\textbf {\bibinfo {volume} {88}},\ \bibinfo {pages} {235113} (\bibinfo
  {year} {2013})}\BibitemShut {NoStop}%
\bibitem [{\citenamefont {Springborg}\ and\ \citenamefont
  {Kirtman}(2008)}]{springborg}%
  \BibitemOpen
  \bibfield  {author} {\bibinfo {author} {\bibfnamefont {Michael}\ \bibnamefont
  {Springborg}}\ and\ \bibinfo {author} {\bibfnamefont {Bernard}\ \bibnamefont
  {Kirtman}},\ }\bibfield  {title} {\enquote {\bibinfo {title} {Analysis of
  vector potential approach for calculating linear and nonlinear responses of
  infinite periodic systems to a finite static external electric field},}\
  }\href@noop {} {\bibfield  {journal} {\bibinfo  {journal} {Phys. Rev. B}\
  }\textbf {\bibinfo {volume} {77}},\ \bibinfo {pages} {045102} (\bibinfo
  {year} {2008})}\BibitemShut {NoStop}%
\bibitem [{\citenamefont {Attaccalite}\ \emph {et~al.}(2011)\citenamefont
  {Attaccalite}, \citenamefont {Gr\"uning},\ and\ \citenamefont
  {Marini}}]{attaccalite}%
  \BibitemOpen
  \bibfield  {author} {\bibinfo {author} {\bibfnamefont {C.}~\bibnamefont
  {Attaccalite}}, \bibinfo {author} {\bibfnamefont {M.}~\bibnamefont
  {Gr\"uning}}, \ and\ \bibinfo {author} {\bibfnamefont {A.}~\bibnamefont
  {Marini}},\ }\bibfield  {title} {\enquote {\bibinfo {title} {Real-time
  approach to the optical properties of solids and nanostructures:
  Time-dependent bethe-salpeter equation},}\ }\href@noop {} {\bibfield
  {journal} {\bibinfo  {journal} {Phys. Rev. B}\ }\textbf {\bibinfo {volume}
  {84}},\ \bibinfo {pages} {245110} (\bibinfo {year} {2011})}\BibitemShut
  {NoStop}%
\bibitem [{\citenamefont {Adler}(1962)}]{PhysRev.126.413}%
  \BibitemOpen
  \bibfield  {author} {\bibinfo {author} {\bibfnamefont {Stephen~L.}\
  \bibnamefont {Adler}},\ }\bibfield  {title} {\enquote {\bibinfo {title}
  {Quantum theory of the dielectric constant in real solids},}\ }\href@noop {}
  {\bibfield  {journal} {\bibinfo  {journal} {Phys. Rev.}\ }\textbf {\bibinfo
  {volume} {126}},\ \bibinfo {pages} {413--420} (\bibinfo {year}
  {1962})}\BibitemShut {NoStop}%
\bibitem [{\citenamefont {Koonin}\ and\ \citenamefont {C.}(2008)}]{koonin90}%
  \BibitemOpen
  \bibinfo {editor} {\bibfnamefont {Steven~E.}\ \bibnamefont {Koonin}}\ and\
  \bibinfo {editor} {\bibfnamefont {Meredith~Dawn}\ \bibnamefont {C.}},\ eds.,\
  \href@noop {} {\emph {\bibinfo {title} {Computational Physics: Fortran
  Version}}}\ (\bibinfo  {publisher} {Perseus Books},\ \bibinfo {year}
  {2008})\BibitemShut {NoStop}%
\bibitem [{\citenamefont {Smalley}\ \emph {et~al.}(2003)\citenamefont
  {Smalley}, \citenamefont {Dresselhaus}, \citenamefont {Dresselhaus},\ and\
  \citenamefont {Avouris}}]{smalley2003carbon}%
  \BibitemOpen
  \bibfield  {author} {\bibinfo {author} {\bibfnamefont {Richard~E}\
  \bibnamefont {Smalley}}, \bibinfo {author} {\bibfnamefont {Mildred~S}\
  \bibnamefont {Dresselhaus}}, \bibinfo {author} {\bibfnamefont {Gene}\
  \bibnamefont {Dresselhaus}}, \ and\ \bibinfo {author} {\bibfnamefont
  {Phaedon}\ \bibnamefont {Avouris}},\ }\href@noop {} {\emph {\bibinfo {title}
  {Carbon nanotubes: synthesis, structure, properties, and applications}}},\
  Vol.~\bibinfo {volume} {80}\ (\bibinfo  {publisher} {Springer Science \&
  Business Media},\ \bibinfo {year} {2003})\BibitemShut {NoStop}%
\bibitem [{\citenamefont {Nemilentsau}\ \emph
  {et~al.}(2006{\natexlab{b}})\citenamefont {Nemilentsau}, \citenamefont
  {Slepyan}, \citenamefont {Khrutchinskii},\ and\ \citenamefont
  {Maksimenko}}]{Nemilentsau20062246}%
  \BibitemOpen
  \bibfield  {author} {\bibinfo {author} {\bibfnamefont {A.M.}\ \bibnamefont
  {Nemilentsau}}, \bibinfo {author} {\bibfnamefont {G.Ya.}\ \bibnamefont
  {Slepyan}}, \bibinfo {author} {\bibfnamefont {A.A.}\ \bibnamefont
  {Khrutchinskii}}, \ and\ \bibinfo {author} {\bibfnamefont {S.A.}\
  \bibnamefont {Maksimenko}},\ }\bibfield  {title} {\enquote {\bibinfo {title}
  {Third-order optical nonlinearity in single-wall carbon nanotubes},}\ }\href
  {\doibase http://dx.doi.org/10.1016/j.carbon.2006.02.035} {\bibfield
  {journal} {\bibinfo  {journal} {Carbon}\ }\textbf {\bibinfo {volume} {44}},\
  \bibinfo {pages} {2246 -- 2253} (\bibinfo {year}
  {2006}{\natexlab{b}})}\BibitemShut {NoStop}%
\bibitem [{\citenamefont {Xu}\ and\ \citenamefont {Xiong}(2004)}]{xu2004third}%
  \BibitemOpen
  \bibfield  {author} {\bibinfo {author} {\bibfnamefont {Yong}\ \bibnamefont
  {Xu}}\ and\ \bibinfo {author} {\bibfnamefont {Guiguang}\ \bibnamefont
  {Xiong}},\ }\bibfield  {title} {\enquote {\bibinfo {title} {Third-order
  optical nonlinearity of semiconductor carbon nanotubes for third harmonic
  generation},}\ }\href@noop {} {\bibfield  {journal} {\bibinfo  {journal}
  {Chem. Phys. Lett.}\ }\textbf {\bibinfo {volume} {388}},\ \bibinfo {pages}
  {330--336} (\bibinfo {year} {2004})}\BibitemShut {NoStop}%
\bibitem [{\citenamefont {Saarinen}(2002)}]{saarinen2002sum}%
  \BibitemOpen
  \bibfield  {author} {\bibinfo {author} {\bibfnamefont {JJ}~\bibnamefont
  {Saarinen}},\ }\bibfield  {title} {\enquote {\bibinfo {title} {Sum rules for
  arbitrary-order harmonic generation susceptibilities},}\ }\href@noop {}
  {\bibfield  {journal} {\bibinfo  {journal} {Eur. Phys. J. B}\ }\textbf
  {\bibinfo {volume} {30}},\ \bibinfo {pages} {551--557} (\bibinfo {year}
  {2002})}\BibitemShut {NoStop}%
\bibitem [{\citenamefont {Son}\ \emph {et~al.}(2006)\citenamefont {Son},
  \citenamefont {Cohen},\ and\ \citenamefont {Louie}}]{son2006energy}%
  \BibitemOpen
  \bibfield  {author} {\bibinfo {author} {\bibfnamefont {Young-Woo}\
  \bibnamefont {Son}}, \bibinfo {author} {\bibfnamefont {Marvin~L}\
  \bibnamefont {Cohen}}, \ and\ \bibinfo {author} {\bibfnamefont {Steven~G}\
  \bibnamefont {Louie}},\ }\bibfield  {title} {\enquote {\bibinfo {title}
  {Energy gaps in graphene nanoribbons},}\ }\href@noop {} {\bibfield  {journal}
  {\bibinfo  {journal} {Physical review letters}\ }\textbf {\bibinfo {volume}
  {97}},\ \bibinfo {pages} {216803} (\bibinfo {year} {2006})}\BibitemShut
  {NoStop}%
\bibitem [{\citenamefont {Prezzi}\ \emph {et~al.}(2011)\citenamefont {Prezzi},
  \citenamefont {Varsano}, \citenamefont {Ruini},\ and\ \citenamefont
  {Molinari}}]{PhysRevB.84.041401}%
  \BibitemOpen
  \bibfield  {author} {\bibinfo {author} {\bibfnamefont {Deborah}\ \bibnamefont
  {Prezzi}}, \bibinfo {author} {\bibfnamefont {Daniele}\ \bibnamefont
  {Varsano}}, \bibinfo {author} {\bibfnamefont {Alice}\ \bibnamefont {Ruini}},
  \ and\ \bibinfo {author} {\bibfnamefont {Elisa}\ \bibnamefont {Molinari}},\
  }\bibfield  {title} {\enquote {\bibinfo {title} {Quantum dot states and
  optical excitations of edge-modulated graphene nanoribbons},}\ }\href
  {\doibase 10.1103/PhysRevB.84.041401} {\bibfield  {journal} {\bibinfo
  {journal} {Phys. Rev. B}\ }\textbf {\bibinfo {volume} {84}},\ \bibinfo
  {pages} {041401} (\bibinfo {year} {2011})}\BibitemShut {NoStop}%
\bibitem [{\citenamefont {Maeda}\ \emph
  {et~al.}(2005{\natexlab{b}})\citenamefont {Maeda}, \citenamefont {Matsumoto},
  \citenamefont {Kishida}, \citenamefont {Takenobu}, \citenamefont {Iwasa},
  \citenamefont {Shiraishi}, \citenamefont {Ata},\ and\ \citenamefont
  {Okamoto}}]{PhysRevLett.94.047404}%
  \BibitemOpen
  \bibfield  {author} {\bibinfo {author} {\bibfnamefont {A.}~\bibnamefont
  {Maeda}}, \bibinfo {author} {\bibfnamefont {S.}~\bibnamefont {Matsumoto}},
  \bibinfo {author} {\bibfnamefont {H.}~\bibnamefont {Kishida}}, \bibinfo
  {author} {\bibfnamefont {T.}~\bibnamefont {Takenobu}}, \bibinfo {author}
  {\bibfnamefont {Y.}~\bibnamefont {Iwasa}}, \bibinfo {author} {\bibfnamefont
  {M.}~\bibnamefont {Shiraishi}}, \bibinfo {author} {\bibfnamefont
  {M.}~\bibnamefont {Ata}}, \ and\ \bibinfo {author} {\bibfnamefont
  {H.}~\bibnamefont {Okamoto}},\ }\bibfield  {title} {\enquote {\bibinfo
  {title} {Large optical nonlinearity of semiconducting single-walled carbon
  nanotubes under resonant excitations},}\ }\href {\doibase
  10.1103/PhysRevLett.94.047404} {\bibfield  {journal} {\bibinfo  {journal}
  {Phys. Rev. Lett.}\ }\textbf {\bibinfo {volume} {94}},\ \bibinfo {pages}
  {047404} (\bibinfo {year} {2005}{\natexlab{b}})}\BibitemShut {NoStop}%
\bibitem [{\citenamefont {Pedersen}\ and\ \citenamefont
  {Cornean}(2007)}]{pedersen2007optical}%
  \BibitemOpen
  \bibfield  {author} {\bibinfo {author} {\bibfnamefont {Thomas~G}\
  \bibnamefont {Pedersen}}\ and\ \bibinfo {author} {\bibfnamefont {Horia~D}\
  \bibnamefont {Cornean}},\ }\bibfield  {title} {\enquote {\bibinfo {title}
  {Optical second harmonic generation from wannier excitons},}\ }\href@noop {}
  {\bibfield  {journal} {\bibinfo  {journal} {EPL (Europhys. Lett.)}\ }\textbf
  {\bibinfo {volume} {78}},\ \bibinfo {pages} {27005} (\bibinfo {year}
  {2007})}\BibitemShut {NoStop}%
\bibitem [{\citenamefont {Gr\"uning}\ \emph {et~al.}(2016)\citenamefont
  {Gr\"uning}, \citenamefont {Sangalli},\ and\ \citenamefont
  {Attaccalite}}]{gruning2016dielectrics}%
  \BibitemOpen
  \bibfield  {author} {\bibinfo {author} {\bibfnamefont {M.}~\bibnamefont
  {Gr\"uning}}, \bibinfo {author} {\bibfnamefont {D.}~\bibnamefont {Sangalli}},
  \ and\ \bibinfo {author} {\bibfnamefont {C.}~\bibnamefont {Attaccalite}},\
  }\bibfield  {title} {\enquote {\bibinfo {title} {Dielectrics in a
  time-dependent electric field: A real-time approach based on
  density-polarization functional theory},}\ }\href {\doibase
  10.1103/PhysRevB.94.035149} {\bibfield  {journal} {\bibinfo  {journal} {Phys.
  Rev. B}\ }\textbf {\bibinfo {volume} {94}},\ \bibinfo {pages} {035149}
  (\bibinfo {year} {2016})}\BibitemShut {NoStop}%
\bibitem [{\citenamefont {Luppi}\ \emph {et~al.}(2010)\citenamefont {Luppi},
  \citenamefont {H\"ubener},\ and\ \citenamefont
  {V\'eniard}}]{PhysRevB.82.235201}%
  \BibitemOpen
  \bibfield  {author} {\bibinfo {author} {\bibfnamefont {Eleonora}\
  \bibnamefont {Luppi}}, \bibinfo {author} {\bibfnamefont {Hannes}\
  \bibnamefont {H\"ubener}}, \ and\ \bibinfo {author} {\bibfnamefont
  {Val\'erie}\ \bibnamefont {V\'eniard}},\ }\bibfield  {title} {\enquote
  {\bibinfo {title} {Ab initio second-order nonlinear optics in solids:
  Second-harmonic generation spectroscopy from time-dependent
  density-functional theory},}\ }\href@noop {} {\bibfield  {journal} {\bibinfo
  {journal} {Phys. Rev. B}\ }\textbf {\bibinfo {volume} {82}},\ \bibinfo
  {pages} {235201} (\bibinfo {year} {2010})}\BibitemShut {NoStop}%
\bibitem [{\citenamefont {Kirtman}\ \emph {et~al.}(2015)\citenamefont
  {Kirtman}, \citenamefont {Springborg}, \citenamefont {Rérat}, \citenamefont
  {Ferrero}, \citenamefont {Lacivita}, \citenamefont {Orlando},\ and\
  \citenamefont {Dovesi}}]{newcrystal}%
  \BibitemOpen
  \bibfield  {author} {\bibinfo {author} {\bibfnamefont {Bernard}\ \bibnamefont
  {Kirtman}}, \bibinfo {author} {\bibfnamefont {Michael}\ \bibnamefont
  {Springborg}}, \bibinfo {author} {\bibfnamefont {Michel}\ \bibnamefont
  {Rérat}}, \bibinfo {author} {\bibfnamefont {Mauro}\ \bibnamefont {Ferrero}},
  \bibinfo {author} {\bibfnamefont {Valentina}\ \bibnamefont {Lacivita}},
  \bibinfo {author} {\bibfnamefont {Roberto}\ \bibnamefont {Orlando}}, \ and\
  \bibinfo {author} {\bibfnamefont {Roberto}\ \bibnamefont {Dovesi}},\
  }\bibfield  {title} {\enquote {\bibinfo {title} {The linear and nonlinear
  response of infinite periodic systems to static and/or dynamic electric
  fields. implementation in crystal code},}\ }\href {\doibase
  http://dx.doi.org/10.1063/1.4906650} {\bibfield  {journal} {\bibinfo
  {journal} {AIP Conference Proceedings}\ }\textbf {\bibinfo {volume} {1642}},\
  \bibinfo {pages} {193--196} (\bibinfo {year} {2015})}\BibitemShut {NoStop}%
\end{thebibliography}


\end{document}